\documentclass[10pt,twocolumn,letterpaper]{article}

\usepackage{iccv}
\usepackage{times}
\usepackage{epsfig}
\usepackage{graphicx}
\usepackage{amsmath}
\usepackage{amssymb}

\usepackage{soul} 
\usepackage{bm}
\usepackage{makecell}
\usepackage{multirow}
\usepackage{bigstrut}
\usepackage{array}
\newcolumntype{L}[1]{>{\raggedright\let\newline\\\arraybackslash\hspace{0pt}}m{#1}}
\newcolumntype{C}[1]{>{\centering\let\newline\\\arraybackslash\hspace{0pt}}m{#1}}
\newcolumntype{R}[1]{>{\raggedleft\let\newline\\\arraybackslash\hspace{0pt}}m{#1}}

\usepackage{graphicx}
\usepackage[justification=centering,%
labelformat=simple]{subfig}

\usepackage{enumitem}

\graphicspath{{figures/}{subpictures/}{figures/material_figures/}}

\makeatletter
\newcommand\datasize{\@setfontsize\normalize{6.32pt}{7.63pt}} 
\makeatother

\makeatletter
\def\CTEX@fs@eight{5.02}   \def\CTEX@fs@eightskip{6.02}   
\def\CTEX@fs@seven{5.52}   \def\CTEX@fs@sevenskip{6.62}   
\def\CTEX@fs@ssix{6.52}    \def\CTEX@fs@ssixskip{7.83}    
\def\CTEX@fs@six{7.53}     \def\CTEX@fs@sixskip{9.03}     
\def\CTEX@fs@sfive{9.03}   \def\CTEX@fs@sfiveskip{10.84}  
\def\CTEX@fs@five{10.54}   \def\CTEX@fs@fiveskip{12.65}   
\def\CTEX@fs@sfour{12.05}  \def\CTEX@fs@sfourskip{14.45}  
\def\CTEX@fs@four{14.05}   \def\CTEX@fs@fourskip{16.86}   
\def\CTEX@fs@sthree{15.06} \def\CTEX@fs@sthreeskip{18.07} 
\def\CTEX@fs@three{16.06}  \def\CTEX@fs@threeskip{19.27}  
\def\CTEX@fs@stwo{18.07}   \def\CTEX@fs@stwoskip{21.68}   
\def\CTEX@fs@two{22.08}    \def\CTEX@fs@twoskip{26.50}    
\def\CTEX@fs@sone{24.09}   \def\CTEX@fs@soneskip{28.91}   
\def\CTEX@fs@one{26.10}    \def\CTEX@fs@oneskip{31.32}    
\def\CTEX@fs@szero{36.14}  \def\CTEX@fs@szeroskip{43.36}  
\def\CTEX@fs@zero{42.16}   \def\CTEX@fs@zeroskip{50.59}   
\DeclareMathSizes{\CTEX@fs@eight}{\CTEX@fs@eight}{5}{5}
\DeclareMathSizes{\CTEX@fs@seven}{\CTEX@fs@seven}{5}{5}
\DeclareMathSizes{\CTEX@fs@ssix}{\CTEX@fs@ssix}{5}{5}
\DeclareMathSizes{\CTEX@fs@six}{\CTEX@fs@six}{5}{5}
\DeclareMathSizes{\CTEX@fs@sfive}{\CTEX@fs@sfive}{6}{5}
\DeclareMathSizes{\CTEX@fs@five}{\CTEX@fs@five}{7}{5}
\DeclareMathSizes{\CTEX@fs@sfour}{\CTEX@fs@sfour}{8}{6}
\DeclareMathSizes{\CTEX@fs@four}
{\CTEX@fs@four}{\CTEX@fs@five}{\CTEX@fs@six}
\DeclareMathSizes{\CTEX@fs@sthree}
{\CTEX@fs@sthree}{\CTEX@fs@sfour}{\CTEX@fs@sfive}
\DeclareMathSizes{\CTEX@fs@three}
{\CTEX@fs@three}{\CTEX@fs@four}{\CTEX@fs@five}
\DeclareMathSizes{\CTEX@fs@stwo}
{\CTEX@fs@stwo}{\CTEX@fs@sthree}{\CTEX@fs@sfour}
\DeclareMathSizes{\CTEX@fs@two}
{\CTEX@fs@two}{\CTEX@fs@three}{\CTEX@fs@four}
\DeclareMathSizes{\CTEX@fs@sone}
{\CTEX@fs@sone}{\CTEX@fs@stwo}{\CTEX@fs@sthree}
\DeclareMathSizes{\CTEX@fs@one}
{\CTEX@fs@one}{\CTEX@fs@two}{\CTEX@fs@three}
\DeclareMathSizes{\CTEX@fs@szero}
{\CTEX@fs@szero}{\CTEX@fs@sone}{\CTEX@fs@stwo}
\DeclareMathSizes{\CTEX@fs@zero}
{\CTEX@fs@zero}{\CTEX@fs@one}{\CTEX@fs@two}
\def\CTEX@zihao{}
\DeclareRobustCommand*\zihao[1]{\def\CTEX@zihao{#1}%
	\ifnum #11<0%
	\@tempcnta=-#1
	\ifcase\@tempcnta%
	\fontsize\CTEX@fs@szero\CTEX@fs@szeroskip%
	\or \fontsize\CTEX@fs@sone\CTEX@fs@soneskip%
	\or \fontsize\CTEX@fs@stwo\CTEX@fs@stwoskip%
	\or \fontsize\CTEX@fs@sthree\CTEX@fs@sthreeskip%
	\or \fontsize\CTEX@fs@sfour\CTEX@fs@sfourskip%
	\or \fontsize\CTEX@fs@sfive\CTEX@fs@sfiveskip%
	\or \fontsize\CTEX@fs@ssix\CTEX@fs@ssixskip%
	\else \PackageError{components}{%
		Undefined Chinese font size in command \protect\zihao}{%
		The old font size is used if you continue.}%
	\fi%
	\else%
	\@tempcnta=#1
	\ifcase\@tempcnta%
	\fontsize\CTEX@fs@zero\CTEX@fs@zeroskip%
	\or \fontsize\CTEX@fs@one\CTEX@fs@oneskip%
	\or \fontsize\CTEX@fs@two\CTEX@fs@twoskip%
	\or \fontsize\CTEX@fs@three\CTEX@fs@threeskip%
	\or \fontsize\CTEX@fs@four\CTEX@fs@fourskip%
	\or \fontsize\CTEX@fs@five\CTEX@fs@fiveskip%
	\or \fontsize\CTEX@fs@six\CTEX@fs@sixskip%
	\or \fontsize\CTEX@fs@seven\CTEX@fs@sevenskip%
	\or \fontsize\CTEX@fs@eight\CTEX@fs@eightskip%
	\else \PackageError{components}{%
		Undefined Chinese font size in command \protect\zihao}{%
		The old font size is used if you continue.}%
	\fi%
	\fi%
	\selectfont\ignorespaces}
\makeatother

\usepackage[pagebackref=true,breaklinks=true,letterpaper=true,colorlinks,bookmarks=false]{hyperref}

\iccvfinalcopy 

\def\iccvPaperID{4892} 

\ificcvfinal\fi
\begin{document}

\title{CRNet: Image Super-Resolution Using A Convolutional \\ Sparse Coding  Inspired Network}

\author{Menglei Zhang, 
	Zhou Liu, 
	Lei Yu\\
	School of Electronic and Information, Wuhan University, China\\
	{\tt\small \{zmlhome, liuzhou, ly.wd\}@whu.edu.cn}
}

\maketitle


\begin{abstract}
	Convolutional Sparse Coding (CSC) has been attracting more and more attention in recent years, for making full use of image global correlation to improve performance on various computer vision applications. However, very few studies focus on solving CSC based image Super-Resolution (SR) problem. As a consequence, there is no significant progress in this area over a period of time. In this paper, we exploit the natural connection between CSC and Convolutional Neural Networks (CNN) to address CSC based image SR. Specifically, Convolutional Iterative Soft Thresholding Algorithm (CISTA) is introduced to solve CSC problem and it can be implemented using CNN architectures. Then we develop a novel CSC based SR framework analogy to the traditional SC based SR methods. Two models inspired by this framework are proposed for pre-/post-upsampling SR, respectively. Compared with recent state-of-the-art SR methods, both of our proposed models show superior performance in terms of both quantitative and qualitative measurements.
\end{abstract}

\vspace*{-.5cm}
\section{Introduction}
Single Image Super-Resolution (SISR), which aims to restore a visually pleasing High-Resolution (HR) image from its Low-Resolution (LR) version, is still a challenging task within computer vision research community \cite{NTIRE2017, NTIRE2018}. Since multiple solutions exist for the mapping from LR to HR space, SISR is highly ill-posed. To regularize the solution of SISR, various priors of natural images have been exploited, especially the current leading learning-based methods \cite{SCN2015, SRCNN2016, REDNet2016, VDSR2016, DRCN2016, DRRN2017, MemNet2017, EDSR2017, CARN2018, DDBPN2018, MSRN2018, RDN2018} are proposed to directly learn the non-linear LR-HR mapping.

By modeling the sparse prior in natural images, the Sparse Coding (SC) based methods for SR \cite{ScSR2008, JianChaoYang2010, IFC2014} with strong theoretical support are widely used owing to their excellent performance. Considering the complexity in images, these methods divide the image into overlapping patches and aim to jointly train two over-complete dictionaries for LR/HR patches. There are usually three steps in these methods' framework. First, overlapping patches are extracted from input image. Then to reconstruct the HR patch, the sparse representation of LR patch can be applied to the HR dictionary with \textit{the assumption that LR/HR patch pair shares similar sparse representation}. The final HR image is produced by aggregating the recovered HR patches.

\begin{figure}[t]
	\centering
	\includegraphics[width=\linewidth, keepaspectratio]{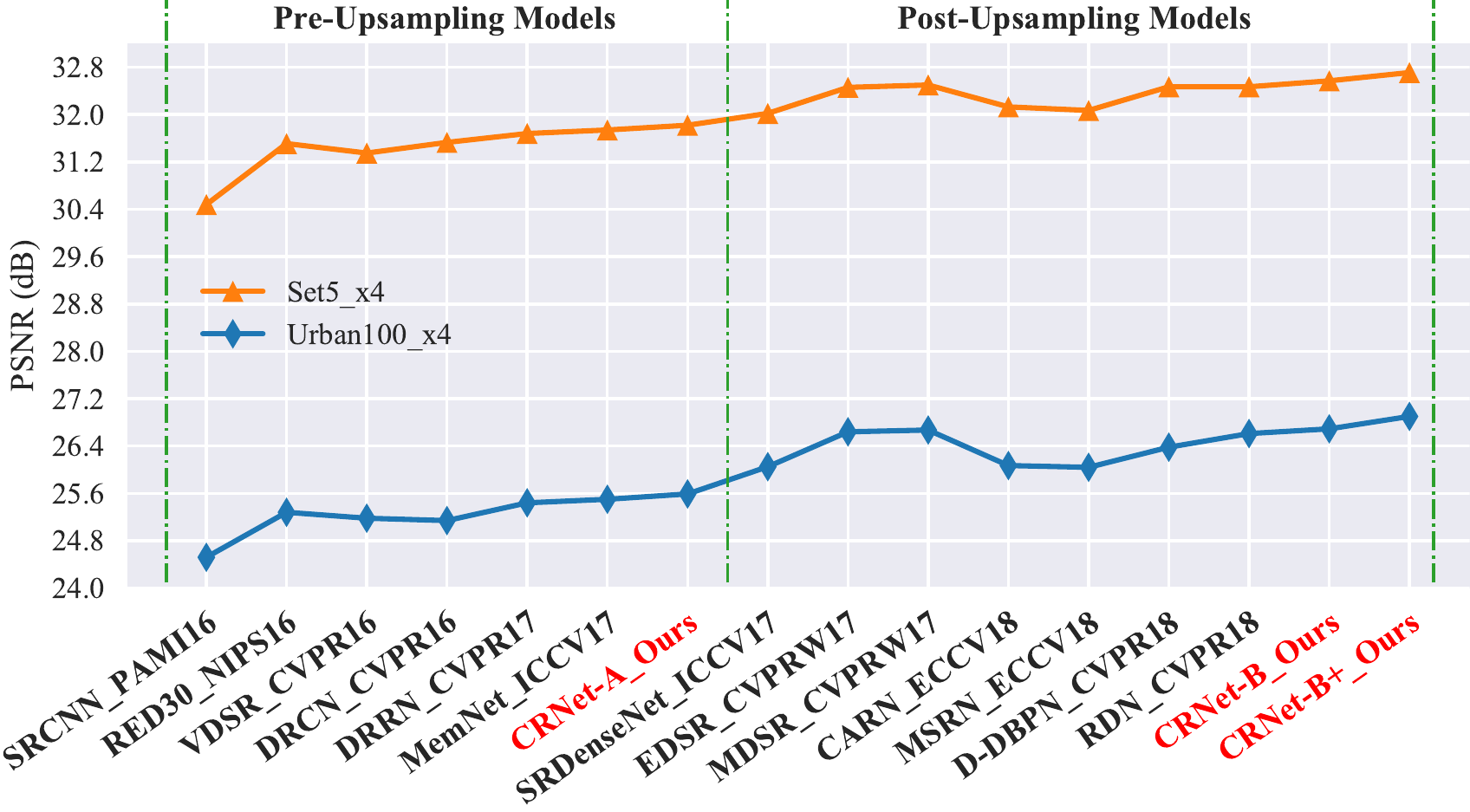}
	\caption{\label{fig:psnr:names} PSNRs of recent state-of-the-arts for scale factor $\times 4$ on Set5 \cite{Set5:2012} and Urban100 \cite{Urban100:2015}. \textcolor{red}{Red} names represent our proposed models.}
	\vspace*{-.5cm}
\end{figure}

Recently, with the development of Deep Learning (DL), many researchers attempt to combine the advantages of DL and SC for image SR.
Dong \etal \cite{SRCNN2016} firstly proposed the seminal CNN model for SR termed as SRCNN, which exploits a shallow convolutional neural network to learn a nonlinear LR-HR mapping in an end-to-end manner and dramatically overshadows conventional methods \cite{JianChaoYang2010, Aplus2014}. 
However, sparse prior is ignored to a large extent in SRCNN for it adopts a generic architecture without considering the domain expertise. To address this issue, Wang \etal \cite{SCN2015} implemented a Sparse Coding based Network (SCN) for image SR, by combining the merits of sparse coding and deep learning,  which fully exploits the approximation of sparse coding learned from the LISTA \cite{LISTA2010} based sub-network. 

\begin{figure}[t]
	\centering
	\includegraphics[width=\linewidth, keepaspectratio]{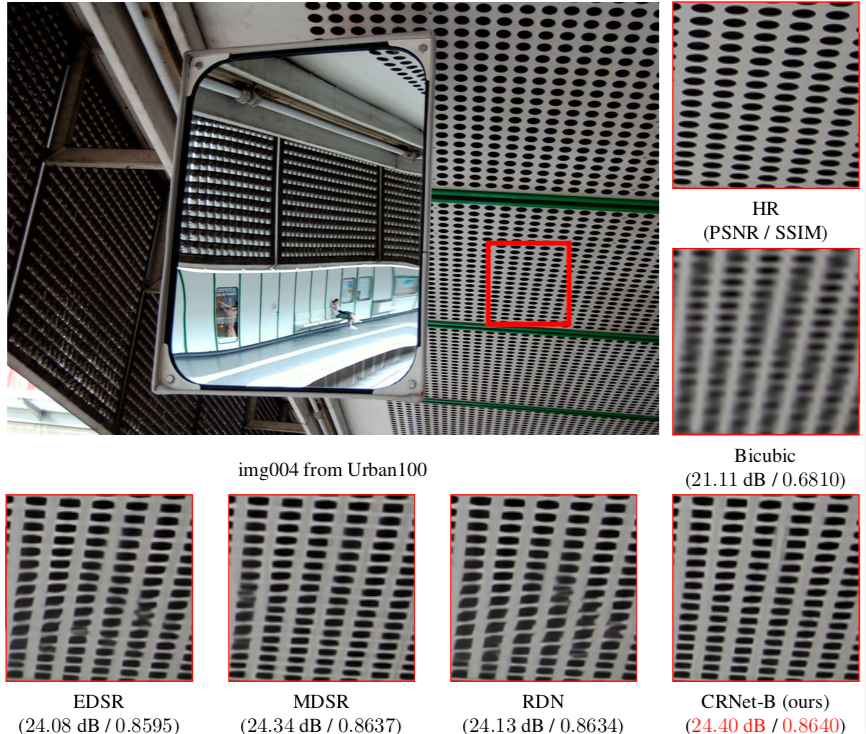}
	\caption{\label{fig:figure2} Visual comparisons between our model CRNet-B and other state-of-the-art methods on ``img004'' from Urban100 \cite{Urban100:2015} for scale factor $\times 4$}
	\vspace*{-.5cm}
\end{figure}

It’s worth to note that most of SC based methods utilize the sparse prior locally \cite{papyan2017working}, \ie, coping with overlapping image patches.  Thus the consistency of pixels in overlapped patches has been ignored \cite{CSC-SR2015, papyan2017working}.
To address this issue, CSC is proposed to serve sparse prior as a global prior \cite{Zeiler:2010kx,papyan2017working, papyan2018theoretical} and it furnishes a way to fill the local-global gap by working directly on the entire image by convolution operation. Consequently, CSC has attained much attention from researchers \cite{Zeiler:2010kx, Bristow:2013fw,Heide:2015hq,CSC-SR2015,LCSC2018,GarciaCardona:2018id}. However, very few studies focus on  the validation of CSC for image SR \cite{CSC-SR2015}, resulting in no work been reported that CSC based image SR can achieve state-of-the-art performance. Can CSC based image SR  show highly competitive results with recent state-of-the-art methods \cite{SRCNN2016, REDNet2016, VDSR2016, DRCN2016, DRRN2017, MemNet2017, SRDenseNet2017, EDSR2017, MSRN2018, RDN2018}? To answer this question, the following issues need to be considered: 

\noindent\textbf{Framework Issue}. Compared with SC based image SR methods \cite{ScSR2008, JianChaoYang2010}, the lack of a unified framework has hindered progress towards improving the performance of CSC based image SR.

\noindent\textbf{Optimization Issue}. The previous CSC based image SR method \cite{CSC-SR2015} contains several steps and they are optimized independently. Hundreds of iterations are required to solve the CSC problem in each step.

\noindent\textbf{Memory Issue}. To solve the CSC problem, ADMM \cite{ADMM2011} is commonly employed \cite{Bristow:2013fw, Wohlberg:2014iy, Heide:2015hq, wohlberg2016boundary, GarciaCardona:2018id}, where the whole training set needs to be loaded in memory. As a consequence, it is not applicable to improve the performance  by enlarging the training set.

\noindent\textbf{Multi-Scale Issue}. Training a single model for multiple scales is difficult for the previous CSC based image SR method~\cite{CSC-SR2015}.

Based on these considerations, in this paper, we attempt to answer the aforementioned question. Specifically, we exploit the advantages of CSC and the powerful learning ability of deep learning to address image SR problem. Moreover, massive theoretical foundations for CSC \cite{papyan2017working, papyan2018theoretical, garcia2018convolutional} make our proposed architectures interpretable and also enable to theoretically analyze our SR performance. In the rest of this paper, we first introduce CISTA, which can be naturally implemented using CNN architectures for solving the  CSC problem. Then we develop a framework for CSC based image SR, which can address the \textbf{Framework Issue}. Subsequently, CRNet-A (CSC and Residual learning based Network) and CRNet-B inspired by this framework are proposed for image SR. They are classified as pre- and post-upsampling models \cite{Wang2019DeepLF} respectively, as the former takes Interpolated LR (ILR) images as input while the latter processes LR images directly. By adopting CNN architectures, \textbf{Optimization Issue} and \textbf{Memory Issue} would be mitigated to some extent.  For \textbf{Multi-Scale Issue}, with the help of the recently introduced scale augmentation \cite{VDSR2016, DRCN2016} or scale-specific multi-path learning \cite{EDSR2017, Wang2019DeepLF} strategies, both of our models are capable of handling multi-scale SR problem effectively, and achieve favorable performance against state-of-the-arts, as shown in Fig.~\ref{fig:psnr:names}. 

The main contributions of this paper include:
\begin{itemize}[noitemsep,topsep=0pt]
	\item We introduce CISTA, which can be naturally implemented using CNN architectures for solving the CSC problem.
	\item A novel framework for CSC based image SR is developed. Two models, CRNet-A and CRNet-B, inspired by this framework are proposed for image SR.
	\item Experimental results demonstrate our proposed models outperform the previous CSC based image SR method \cite{CSC-SR2015} by a large margin and show superior performance against recent state-of-the-arts, \eg, EDSR/MDSR \cite{EDSR2017}, RDN \cite{RDN2018}, as depicted in Fig.~\ref{fig:figure2}.
	\item The differences between our proposed models and several SR models with recursive learning strategy, \eg, DRRN \cite{DRRN2017}, SCN \cite{SCN2015}, DRCN \cite{DRCN2016}, are discussed.
\end{itemize}

\section{Related Work}

\subsection{Sparse Coding for Image Super-Resolution}\label{sub:sparse:coding}

Sparse coding has been widely used in a variety of applications \cite{Zhang:2015kb}. As for SISR, Yang \etal \cite{ScSR2008} proposed a representative Sparse coding based Super-Resolution (ScSR) method. In the  training stage, ScSR attempts to learn the LR/HR overcomplete dictionary pair $\bm{D}_l$/$\bm{D}_h$ jointly by given a group of LR/HR training patch pairs $\bm{x}_l$/$\bm{x}_h$. In the test stage, the HR patch $\bm{x}_h$ is reconstructed from its LR version $\bm{x}_l$ by assuming they share the same sparse code. Specifically, the optimal sparse code is obainted by minimizing the following sparsity-inducing $\ell_1$-norm regularized objective function
\begin{equation}\label{equ:ScSR}
\bm{z}^\ast=\arg \min_{\bm{z}}\left\Vert\bm{x}_l-\bm{D}_l \bm{z}\right\Vert_{2}^{2}+\lambda\Vert\bm{z}\Vert_{1},
\end{equation}
and then the HR patch is obtained by $\bm{x}_h=\bm{D}_h \bm{z}^\ast$. Finally, the HR image can be estimated by aggregating all the reconstructed HR patches. Inspired by ScSR, many SC based SR methods have been proposed by using various constraints on sparse code and dictionary \cite{yang2012coupled, wang2012semi}.

\begin{figure}
	\centering
	\includegraphics[width=\linewidth, keepaspectratio] {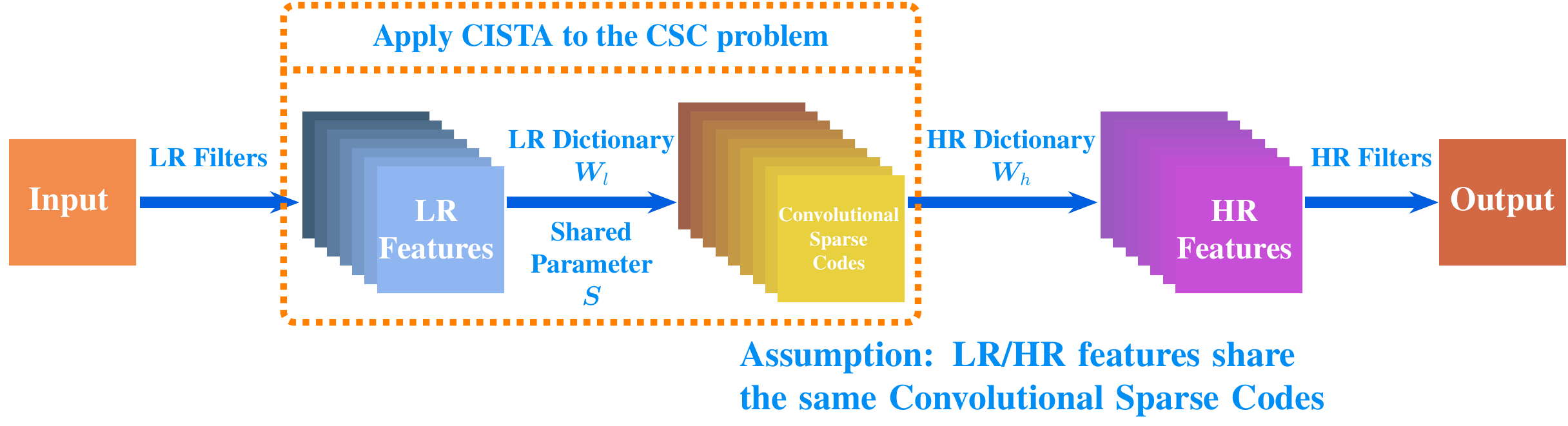}
	\vspace*{-.6cm}
	\caption{\label{fig:flowchar} Our framework for CSC based image SR. }
	\vspace*{-.4cm}
\end{figure}

%

\subsection{Convolutional Sparse Coding for Image Super-Resolution}

Traditional SC based SR algorithms usually process images in a patch based manner to reduce the burden of modeling and computation, resulting in the inconsistency problem \cite{papyan2017working}. As a special case of SC, CSC is inherently suitable for this issue \cite{Zeiler:2010kx}. CSC is proposed to avoid the inconsistency problem by representing the whole image directly. Specifically, an image $\bm{y}\in\mathbb{R}^{n_r\times n_c}$ can be represented as the summation of $m$ feature maps $\bm{z}_i\in\mathbb{R}^{n_r\times n_c}$ convolved with the corresponding filters $\bm{f}_i\in\mathbb{R}^{s\times s}$: $\bm{y} = \sum _ { i = 1 } ^ { m } \bm { f } _ { i } \otimes \bm { z } _ { i }$, where $ \otimes $ is the convolution operation. 

Gu \etal \cite{CSC-SR2015} proposed the CSC-SR method and revealed the potential of CSC for image SR. In \cite{CSC-SR2015}, CSC-SR requires to solve the following CSC based optimization problem in both the training and testing phase:

\begin{equation}\label{equ:csc}
\min _ {\bm{f},\bm{z} } \frac {1}{2} \left\| \bm { y } - \sum _ { i = 1 }  ^ { m } \bm {f} _ {i } \otimes \bm { z } _ { i } \right\| _ { 2 } ^ { 2 } + \lambda \sum _ { i = 1 } ^ { m } \left\| \bm { z } _ { i } \right\| _ { 1 }.
\end{equation}
\cite{CSC-SR2015} solves this problem by alternatively optimizing the $\bm{z}$ and $\bm{f}$ subproblems \cite{Wohlberg:2014iy}. The $\bm{z}$ subproblem is a standard CSC problem. Hundreds of iterations are required to solve the CSC problem and the  aforementioned \textbf{Optimization Issue} and \textbf{Memory Issue} cannot be completely avoided.
%
Inspired by the success of deep learning based sparse coding \cite{LISTA2010}, we exploit the natural connection between CSC and CNN to solve the CSC problem efficiently.

\section{CISTA for solving CSC problem}

\begin{figure}
	%
	\centering
	\renewcommand{\thesubfigure}{(a)} 
	\vspace*{-.2cm}
	\subfloat[$K = 1$]{\label{sub:demo:k1}\includegraphics[%
		width=0.466\linewidth, keepaspectratio]{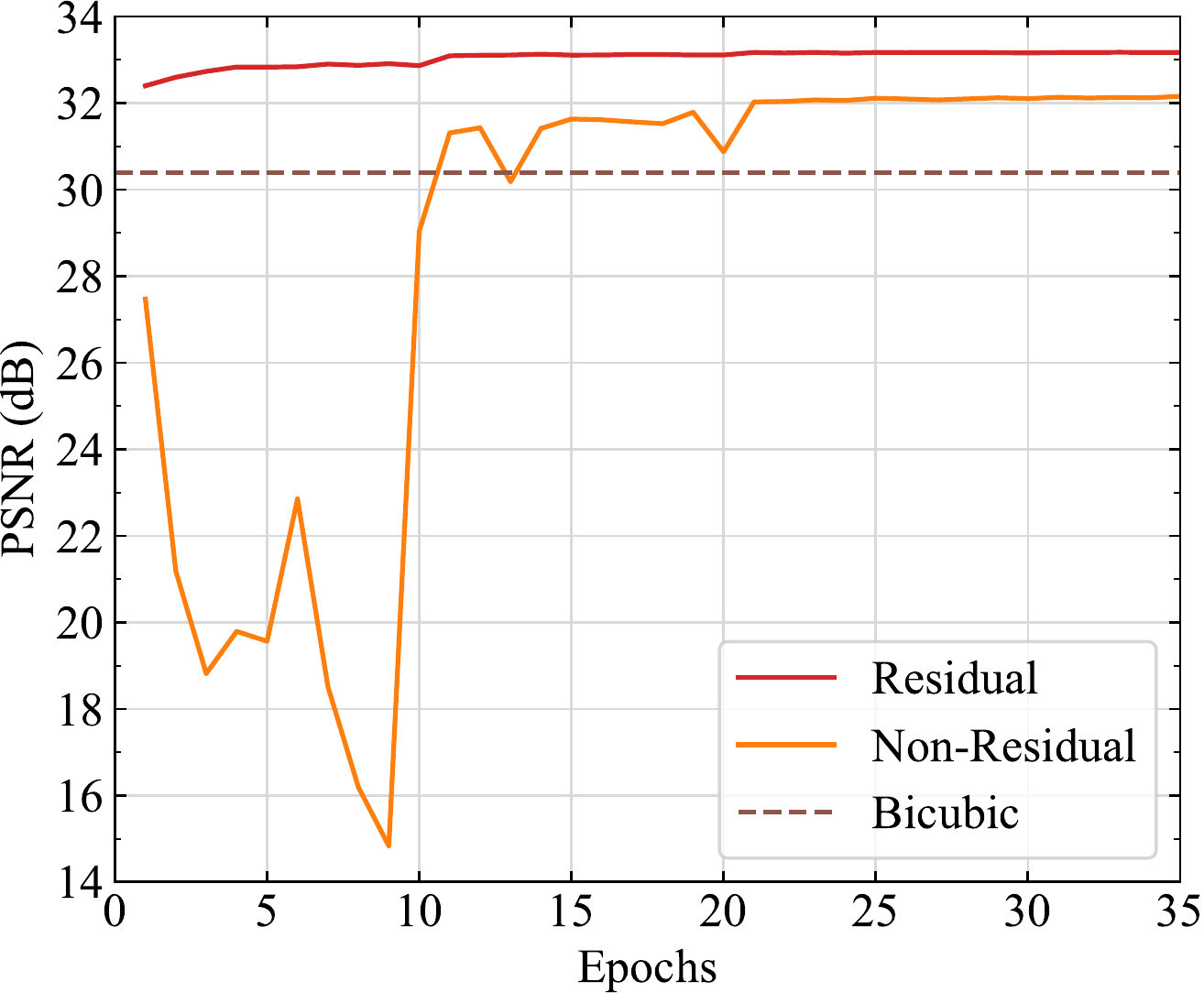}}\hspace*{.03\linewidth}
	\renewcommand{\thesubfigure}{(b)}
	\subfloat[$K = 5$]{\label{sub:ndemo:k5}\includegraphics[%
		width=0.46\linewidth, keepaspectratio]{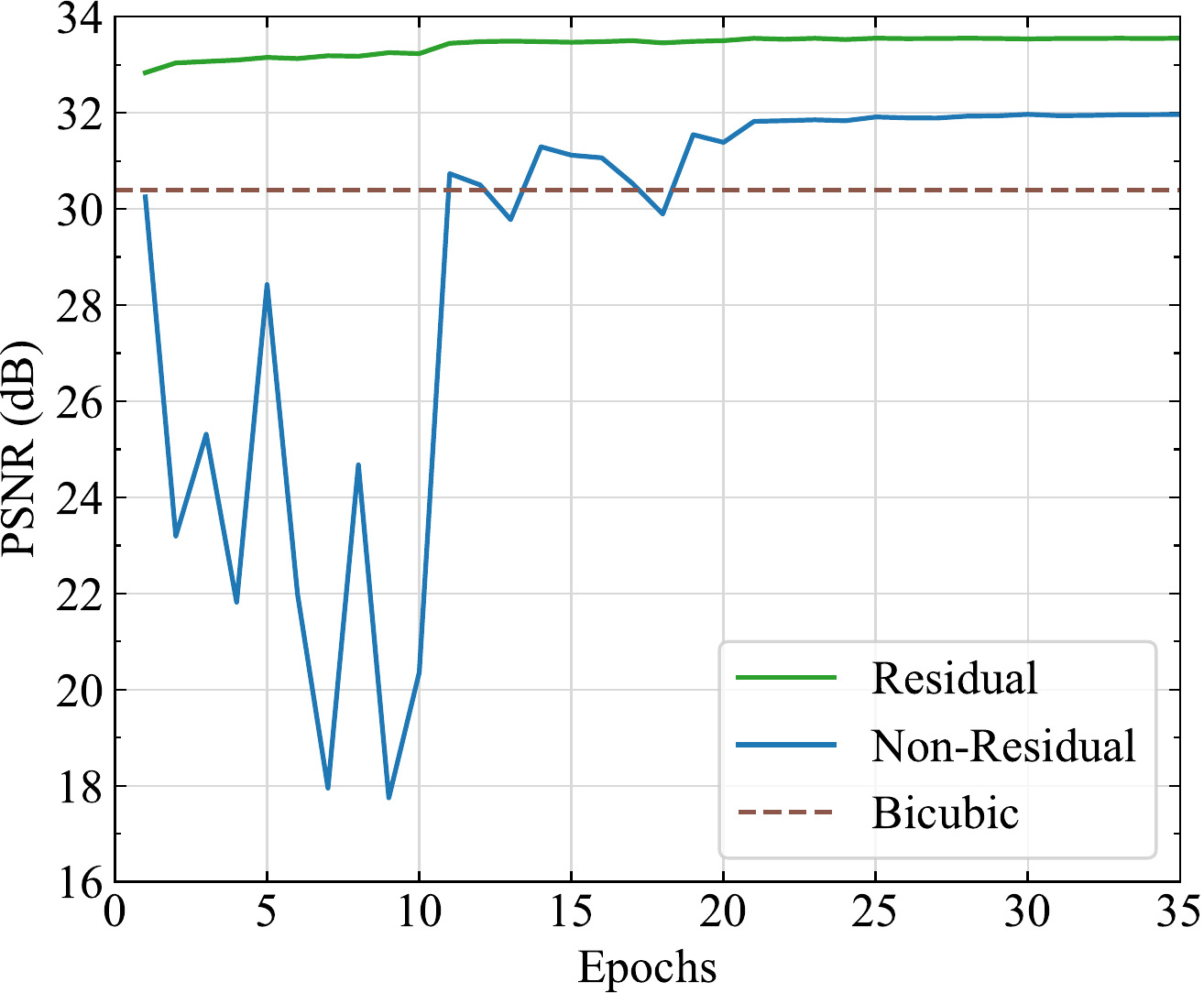}}
	\vspace*{-.2cm}
	\caption{\label{fig:figure3} Performance curve for residual/non-residual networks with different recursions. The tests are conducted on Set5 for scale factor $\times 3$.}
	\vspace*{-.3cm}
\end{figure}

\begin{figure*}
	\centering
	\includegraphics[width=\textwidth, keepaspectratio] {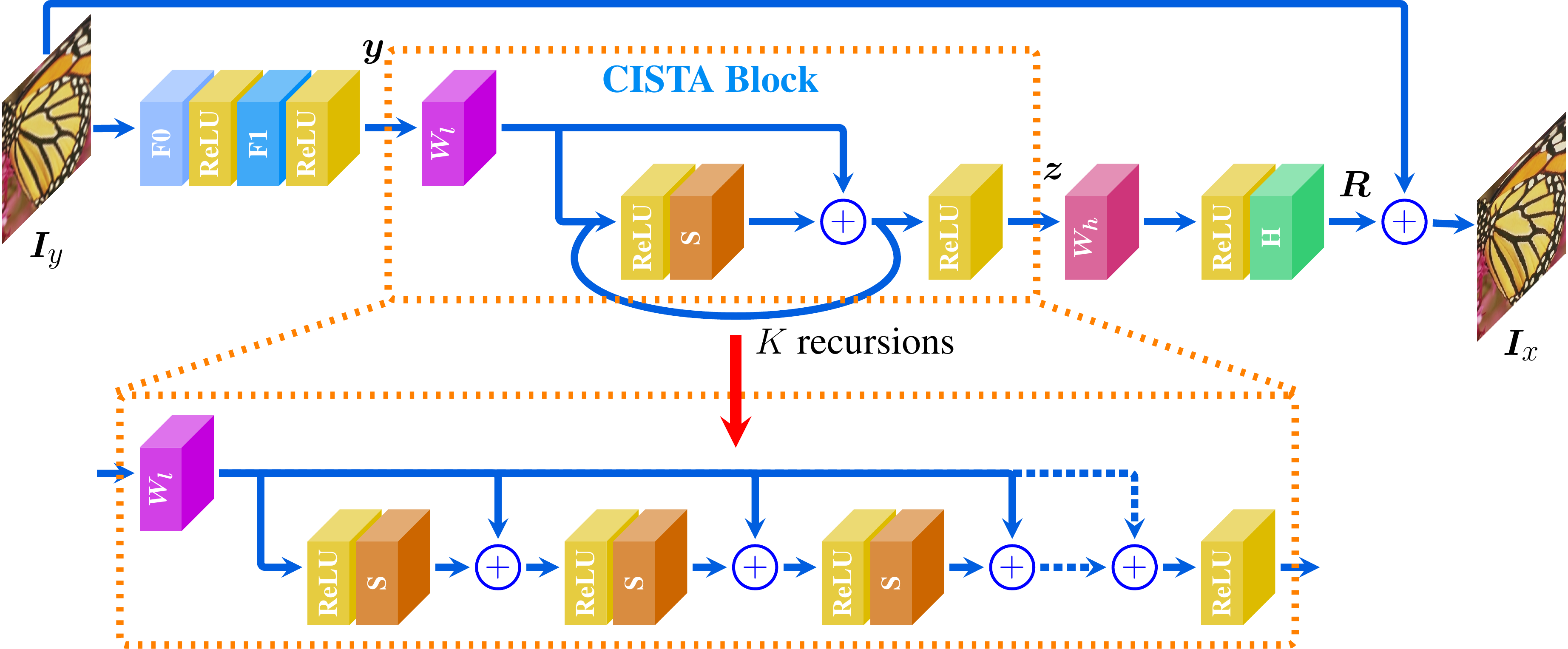}
	\caption{\label{fig:model} The architecture of the pre-upsampling model CRNet-A. The proposed CISTA block with $K$ recursions is surrounded by the dashed box and its unfolded version is shown in the bottom. $\bm{S}$ is shared across every recursion.}
	\vspace*{-.3cm}
\end{figure*}

CSC can be considered as a special case of conventional SC, due to the fact that convolution operation can be replaced with matrix multiplication, so the objective function of CSC can be formulated as:
\begin{equation}\label{equ:rewritten}
\min_{\bm{z}}\left\Vert\bm{y} - \sum_{i=1}^{m}\bm{F}_i\bm{z}_i\right\Vert^2_2+ \lambda\sum_{i=1}^{m}\left\Vert\bm{z}_i\right\Vert_1.
\end{equation}
$\bm{y}, \bm{z}_i$ are in vectorized form and $\bm{F}_i$ is a sparse convolution matrix with the following attributes:
\begin{equation}\label{equ:relation}
\begin{aligned}
\bm{F}_i\bm{z}_i &\equiv \bm{f}_i \otimes \bm{z}_i \\
\bm{F}_i^T\bm{z}_i &\equiv \texttt{flipud}(\texttt{fliplr}(\bm{f}_i)) \otimes \bm{z}_i \\
&\equiv \texttt{flip}(\bm{f}_i) \otimes \bm{z}_i
\end{aligned}
\end{equation}
where $\texttt{fliplr}(\cdot)$ and $\texttt{flipud}(\cdot)$ are following the notations of Zeiler \etal \cite{Zeiler:2010kx}, representing that array is flipped in left/right or up/down direction.

Iterative Soft Thresholding Algorithm (ISTA) \cite{ISTA2004} can be utilized to solve \eqref{equ:rewritten}, at the $k^{th}$ iteration:
\begin{equation}\label{equ:temp}
\bm{z}_{k+1} = h_{\theta}\left(
\bm{z}_k + \frac{1}{L}\bm{F}^T\left(
\bm{y} - \bm{F}\bm{z}_k\right)
\right) 
\end{equation}
where $L$ is the Lipschitz constant, $\bm{F} = [\bm{F}_1, \bm{F}_2, \ldots, \bm{F}_m]$ and $\bm{F}\bm{z} = \sum_{i=1}^{m}\bm{F}_i\bm{z}_i$. Using the relation in \eqref{equ:relation} to replace the matrix multiplication with convolution operator, we can reformulate \eqref{equ:temp} as:
\begin{equation}\label{equ:almost}
\begin{aligned}
\bm{z}_{k+1} 
&= h_{\theta}\left(
\bm{I}\bm{z}_k + \frac{1}{L}\texttt{flip}(\bm{f})\otimes\left(
\bm{y} - \bm{f}\otimes\bm{z}_k\right)
\right)
\end{aligned}
\end{equation}
where $\bm{I}$ is the identity matrix, $\bm{f} = [\bm{f}_1, \bm{f}_2, \ldots, \bm{f}_m]$ and $\texttt{flip}(\bm{f}) = [\texttt{flip}(\bm{f}_1), \texttt{flip}(\bm{f}_2), \ldots, \texttt{flip}(\bm{f}_m)]$. Note that identity matrix $\bm{I}$ is also a sparse convolution matrix, so according to \eqref{equ:relation},  there existing a filter $\bm{n}$ satisfies:
\begin{equation}
\bm{I}\bm{z} = \bm{n}\otimes\bm{z},
\end{equation}
so \eqref{equ:almost} becomes:
\begin{equation}\label{equ:final}
\bm{z}_{k+1} = h_{\theta}\left(
\bm{W}\otimes\bm{y} + \bm{S} \otimes \bm{z}_k
\right),
\end{equation}
where $\bm{W} = \frac{1}{L}\texttt{flip}(\bm{f})$ and $\bm{S} = \bm{n} - \frac{1}{L}\texttt{flip}(\bm{f})\otimes\bm{f}$. Even though \eqref{equ:rewritten} is for a single image with one channel, the extension to multiple channels (for both image and filters) and multiple images is mathematically straightforward. Thus for $\bm{y}\in\mathbb{R}^{b\times c\times n_r\times n_c}$  representing $b$ images of size $n_r\times n_c$ with $c$ channels, \eqref{equ:final} is still true with $\bm{W}\in\mathbb{R}^{m\times c\times s\times s}$ and $\bm{S}\in\mathbb{R}^{m\times m\times s\times s}$.

As for $h_{\bm{\theta}}$, \cite{papyan2017convolutional} reveals two important facts: (1) the expressiveness of the sparsity inspired model is not affected even by restricting the coefficients to be nonnegative; (2) the $ReLU$ \cite{relu2010} activation function and the soft nonnegative thresholding operator are equal, that is:
\begin{equation}\label{equ:relu}
h_{\bm{\theta}}^{+}(\bm{\alpha}) = \max(\bm{\alpha} - \bm{\theta}, 0) = ReLU(\bm{\alpha} - \bm{\theta}).
\end{equation}
We set $\bm{\theta} = \bm{0}$ for simplicity. So the final form of \eqref{equ:final} is:
\begin{equation}\label{equ:cista}
\bm{z}_{k+1} = ReLU\left(
\bm{W}\otimes\bm{y} + \bm{S} \otimes \bm{z}_k
\right).
\end{equation}
One can see that \eqref{equ:cista} is a convolutional form of \eqref{equ:temp}, so we name it as CISTA. It provides the solution of \eqref{equ:rewritten} with theoretical guarantees \cite{ISTA2004}.  Furthermore, this convolutional form can be implemented employing CNN architectures. So $\bm{W}$ and $\bm{S}$ in \eqref{equ:cista} would be trainable.


%

\section{Proposed Method}

\begin{figure*}
	\centering
	\includegraphics[width=\textwidth, keepaspectratio] {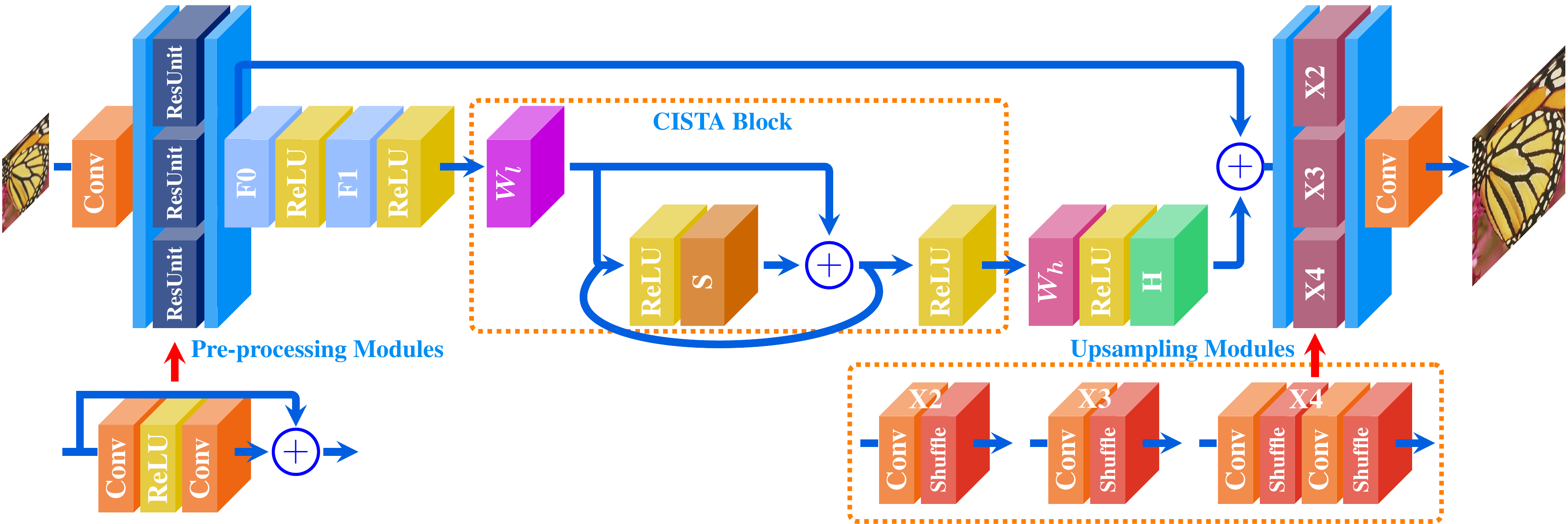}
	\caption{\label{fig:model2} The architecture of the post-upsampling model CRNet-B.}
	\vspace*{-.3cm}
\end{figure*}

In this section, our framework for CSC based image SR is first introduced. And then we implement it using CNN techniques. Since most of image SR methods can be attributed to two frameworks with different upsampling strategies, \ie, pre-upsampling and post-upsampling, we propose two models, CRNet-A for pre-upsampling and CRNet-B for post-upsampling.



\subsection{The framework for CSC based Image SR}

Analogy to sparse coding based SR,  we develop a framework for CSC based image SR. As shown in Fig.~\ref{fig:flowchar}, LR feature maps are extracted from the input LR image using the learned LR filters. Then convolutional sparse codes $\bm{z}$ of LR feature maps are obtained using CISTA with LR dictionary $\bm{W}_l$ and shared parameter $\bm{S}$, as indicated in \eqref{equ:cista}. 
\textbf{Under the assumption that HR feature maps share the same convolutional sparse codes with LR feature maps}, HR feature maps can be recovered by $\bm{W}_h \otimes \bm{z}$. Finally, the HR image is reconstructed by utilizing the learned HR filters.

In this work, we implement this framework  using CNN techniques. However, when combining CSC with CNN, the characteristics of CNN itself must be considered. With more recursions used in CISTA, the network becomes deeper and tends to be bothered by the gradient vanishing/exploding problems. Residual learning \cite{VDSR2016, DRCN2016, DRRN2017} is such a useful tool that not only mitigates these difficulties, but helps network converge faster. In Fig.~\ref{fig:figure3}, residual/non-residual networks with different recursions are compared experimentally and the residual network converges much faster and achieves better performance. Based on these observations, both of our proposed models adopt residual learning.

\subsection{CRNet-A Model for Pre-upsampling}

As shown in Fig.~\ref{fig:model}, CRNet-A takes the ILR image $\bm{I}_y$ with $c$ channels as input, and predicts the output HR image as $\bm{I}_x$. Two convolution layers, $\bm{F}_0\in\mathbb{R}^{n_0\times c\times s\times s}$ consisting of $n_0$ filters of spatial size $c\times s\times s$ and $\bm{F}_1\in\mathbb{R}^{n_0\times n_0\times s\times s}$ containing $n_0$ filters of spatial size $n_0\times s\times s$ are utilized for hierarchical features extraction from ILR image:
\begin{equation}\label{equ:extract}
\bm{y} = ReLU\Big(\bm{F}_1\otimes ReLU(\bm{F}_0\otimes\bm{I}_y)\Big).
\end{equation}

The ILR features are then fed into a CISTA block to learn the convolutional sparse codes. As stated in \eqref{equ:cista}, two convolutional layers $\bm{W}_l\in\mathbb{R}^{m_0\times n_0\times s\times s}$ and $\bm{S}\in\mathbb{R}^{m_0\times m_0\times s\times s}$ are needed:
\begin{equation}
\bm{z}_{k + 1} = ReLU(\bm{W}_l\otimes\bm{y} + \bm{S}\otimes\bm{z}_k),
\end{equation}
where $\bm{z}_0$ is initialized to $ReLU(\bm{W}_l\otimes\bm{y})$. The  convolutional sparse codes $\bm{z}$ are learned after $K$ recursions with $\bm{S}$ shared across every recursion. When the  convolutional sparse codes $\bm{z}$ are obtained, it is then passed through a convolution layer $\bm{W}_h\in\mathbb{R}^{n_0\times m_0\times s\times s}$ to recover the HR feature maps. The last convolution layer $\bm{H}\in\mathbb{R}^{c\times n_0\times s\times s}$ is used as HR filters:
\begin{equation}\label{equ:plus}
\bm{R} = \bm{H}\otimes ReLU(\bm{W}_h \otimes \bm{z} ).
\end{equation}
Note that we pad zeros before all convolution operations to keep all the feature maps to have the same size, which is a common strategy used in a variety of methods \cite{VDSR2016, DRCN2016, DRRN2017}. So the residual image $\bm{R}$ has the same size as the input ILR image $\bm{I}_y$, and the final HR image $\bm{I}_x$ would be reconstructed by: 
\begin{equation}\label{equ:add}
\bm{I}_x = \bm{I}_y + \bm{R}.
\end{equation}
Given $N$ ILR-HR image patch pairs  $\{ \bm{I}_y^{(i)}, \tilde{\bm{I}}_x^{(i)} \}_{i=1}^{N}$ as a training set, our goal is to minimize the following objective function:
\begin{equation}\label{equ:loss}
\mathcal{L}(\bm{\Theta}) = \frac{1}{2N}\sum_{i = 1}^{N}\left\Vert \bm{I}_x^{(i)} - \tilde{\bm{I}}_x^{(i)}\right\Vert^2_2
\end{equation}
where $\bm{\Theta}$ denotes the learnable parameters. The network is optimized using the mini-batch Stochastic Gradient Descent (SGD) with backpropagation~\cite{lecun1998gradient}.


\subsection{CRNet-B Model for Post-upsampling}

We extend CRNet-A to its post-upsampling version to further mine its potential. Notice that most post-upsampling models \cite{SRResNet2017, SRDenseNet2017, MSRN2018, RDN2018} need to train and store many scale-dependent models for various scales without fully using the inter-scale correlation, so we adopt the scale-specific multi-path learning strategy \cite{Wang2019DeepLF} presented in MDSR \cite{EDSR2017} with minor modifications to address this issue. The complete model is shown in Fig.~\ref{fig:model2}. The main branch is our CRNet-A module. The pre-processing modules are used for reducing the variance from input images of different scales and only one residual unit with $3\times 3$ kernels is used in each of the pre-processing module. At the end of CRNet-B, upsampling modules are used for multi-scale reconstruction.

{	
	\begin{table}
		\datasize
		\centering
		\begin{tabular}{|C{.88cm}|C{.6cm}|C{.9cm}|C{.9cm}|C{1.cm}|C{1.3cm}|}
			\hline
			\makecell{Dataset\\[-.13cm]} & \makecell{Scale\\[-.13cm]} & \makecell{Bicubic\\[-.13cm]} &
			\makecell{CSC-SR \\\cite{CSC-SR2015}} &  \makecell{CRNet-B \\ (ours)} & \makecell{Our \\ Improvement}\\\hline\hline\vspace*{.03cm}
			\multirow{3}{*}{Set5} & $\times 2$ & $33.66$ & $36.62$ & $\textcolor{red}{38.13}$ & $\textbf{1.51}$\\
			& $\times 3$ & $30.39$ & $32.65$ & $\textcolor{red}{34.75}$ & $\textbf{2.10}$ \\
			& $\times 4$ & $28.42$ & $30.36$ & $\textcolor{red}{32.57}$ & $\textbf{2.21}$ \\\hline 
		\end{tabular}
		\vspace*{-.1cm}
		\caption{\label{tab:compare:cscsr}Average PSNRs of CSC-SR \cite{CSC-SR2015} and CRNet-B for scale factor $\times2$, $\times3$ and $\times4$ on Set5. The performance gain of our model over CSC-SR is shown in the last column.}
		\vspace*{-.4cm}
	\end{table}
}


\section{Experimental Results}

{	
	\begin{table*}
		\datasize
		\centering
		\begin{tabular}{|C{1.cm}|C{.5cm}|C{1.4cm}|C{1.4cm}|C{1.4cm}|C{1.4cm}|C{1.4cm}|C{1.4cm}|C{1.4cm}|C{1.4cm}|}
			\hline
			\makecell{Dataset\\[-.13cm]} & \makecell{Scale\\[-.13cm]} & \makecell{Bicubic\\[-.13cm]} & \makecell{SRCNN \\ \cite{SRCNN2016}} & \makecell{RED30 \\\cite{REDNet2016}} & \makecell{VDSR \\\cite{VDSR2016}} & \makecell{DRCN \\ \cite{DRCN2016}} & \makecell{DRRN \\ \cite{DRRN2017}} & \makecell{MemNet \\\cite{MemNet2017}} & \makecell{CRNet-A \\ (ours)}\\\hline\hline\vspace*{.03cm}
			\multirow{3}{*}{Set5} & $\times 2$ & $33.66$/$0.9299$ & $36.66$/$0.9542$ & $37.66$/$\textcolor{blue}{0.9599}$ & $37.53$/$0.9587$ & $37.63$/$0.9588$ & $37.74$/$0.9591$ & $\textcolor{blue}{37.78}$/$0.9597$ & $\textcolor{red}{37.79}$/$\textcolor{red}{0.9600}$\\
			& $\times 3$ & $30.39$/$0.8682$ & $32.75$/$0.9090$ & $33.82$/$0.9230$ & $33.66$/$0.9213$ & $33.82$/$0.9226$ & $34.03$/$0.9244$ & $\textcolor{blue}{34.09}$/$\textcolor{blue}{0.9248}$ & $\textcolor{red}{34.11}$/$\textcolor{red}{0.9254}$\\
			& $\times 4$ & $28.42$/$0.8104$ & $30.48$/$0.8628$ & $31.51$/$0.8869$ & $31.35$/$0.8838$ & $31.53$/$0.8854$ & $31.68$/$0.8888$ & $\textcolor{blue}{31.74}$/$\textcolor{blue}{0.8893}$ & $\textcolor{red}{31.82}$/$\textcolor{red}{0.8907}$\\\hline\hline\vspace*{.03cm}
			\multirow{3}{*}{Set14} & $\times 2$ & $30.24$/$0.8688$ & $32.45$/$0.9067$ & $32.94$/$\textcolor{blue}{0.9144}$ & $33.03$/$0.9124$ & $33.04$/$0.9118$ & $33.23$/$0.9136$ & $\textcolor{blue}{33.28}$/$0.9142$ & $\textcolor{red}{33.33}$/$\textcolor{red}{0.9152}$\\
			& $\times 3$ & $27.55$/$0.7742$ & $29.30$/$0.8215$ & $29.61$/$0.8341$ & $29.77$/$0.8314$ & $29.76$/$0.8311$ & $29.96$/$0.8349$ & $\textcolor{red}{30.00}$/$\textcolor{blue}{0.8350}$ & $\textcolor{blue}{29.99}$/$\textcolor{red}{0.8359}$\\
			& $\times 4$ & $26.00$/$0.7027$ & $27.50$/$0.7513$ & $27.86$/$0.7718$ & $28.01$/$0.7674$ & $28.02$/$0.7670$ & $28.21$/$0.7720$ & $\textcolor{blue}{28.26}$/$\textcolor{blue}{0.7723}$ & $\textcolor{red}{28.29}$/$\textcolor{red}{0.7741}$\\\hline\hline\vspace*{.03cm}
			\multirow{3}{*}{B100} & $\times 2$ & $29.56$/$0.8431$ & $31.36$/$0.8879$ & $31.99$/$0.8974$ & $31.90$/$0.8960$ & $31.85$/$0.8942$ & $32.05$/$0.8973$ & $\textcolor{blue}{32.08}$/$\textcolor{blue}{0.8978}$ & $\textcolor{red}{32.09}$/$\textcolor{red}{0.8985}$\\
			& $\times 3$ & $27.21$/$0.7385$ & $28.41$/$0.7863$ & $28.93$/$0.7994$ & $28.82$/$0.7976$ & $28.80$/$0.7963$ & $28.95$/$\textcolor{blue}{0.8004}$ & $\textcolor{blue}{28.96}$/$0.8001$ & $\textcolor{red}{28.99}$/$\textcolor{red}{0.8021}$\\
			& $\times 4$ & $25.96$/$0.6675$ & $26.90$/$0.7101$ & $27.40$/$\textcolor{blue}{0.7290}$ & $27.29$/$0.7251$ & $27.23$/$0.7233$ & $27.38$/$0.7284$ & $\textcolor{blue}{27.40}$/$0.7281$ & $\textcolor{red}{27.44}$/$\textcolor{red}{0.7302}$\\\hline\hline\vspace*{.03cm}
			\multirow{3}{*}{Urban100} & $\times 2$ & $26.88$/$0.8403$ & $29.50$/$0.8946$ & $30.85$/$0.9148$ & $30.76$/$0.9140$ & $30.75$/$0.9133$ & $31.23$/$0.9188$ & $\textcolor{blue}{31.31}$/$\textcolor{blue}{0.9195}$ & $\textcolor{red}{31.36}$/$\textcolor{red}{0.9207}$\\
			& $\times 3$ & $24.46$/$0.7349$ & $26.24$/$0.7989$ & $27.25$/$0.8283$ & $27.14$/$0.8279$ & $27.15$/$0.8276$ & $27.53$/$\textcolor{blue}{0.8378}$ & $\textcolor{blue}{27.56}$/$0.8376$ & $\textcolor{red}{27.64}$/$\textcolor{red}{0.8403}$\\
			& $\times 4$ & $23.14$/$0.6577$ & $24.52$/$0.7221$ & $25.28$/$0.7555$ & $25.18$/$0.7524$ & $25.14$/$0.7510$ & $25.44$/$\textcolor{blue}{0.7638}$ & $\textcolor{blue}{25.50}$/$0.7630$ & $\textcolor{red}{25.59}$/$\textcolor{red}{0.7680}$
			\\\hline
		\end{tabular}
		\vspace*{-.1cm}
		\caption{\label{tab:psnr1}Average PSNR/SSIMs of \textbf{Pre-upsampling} models for scale factor $\times2$, $\times3$ and $\times4$ on datasets Set5, Set14, BSD100 and Urban100. \textcolor{red}{Red} color indicates the best performance and \textcolor{blue}{blue} color indicates the second best performance.}
	\end{table*}
}

{
	
	\begin{table*}
		\datasize
		\centering
		\vspace*{-.3cm}
		\begin{tabular}{|C{1.cm}|C{.5cm}|C{1.4cm}|C{1.4cm}|C{1.4cm}|C{1.4cm}|C{1.4cm}|C{1.4cm}|C{1.4cm}|C{1.4cm}|}
			\hline
			\makecell{Dataset\\[-.13cm]} & \makecell{Scale\\[-.13cm]} & \makecell{SRDenseNet \\ \cite{SRDenseNet2017}} & \makecell{MSRN \\ \cite{MSRN2018}} & \makecell{D-DBPN \\ \cite{DDBPN2018}}& \makecell{EDSR \\ \cite{EDSR2017}} & \makecell{MDSR \\ \cite{EDSR2017}} & \makecell{RDN \\ \cite{RDN2018}}& \makecell{CRNet-B \\ (ours)} & \makecell{CRNet-B+ \\ (ours)}  \\\hline\hline\vspace*{.03cm}
			\multirow{3}{*}{Set5} & $\times 2$ & -/- & $38.08$/$0.9605$ & $38.09$/$0.9600$ & $38.11$/$0.9601$ & $38.11$/$0.9602$ & $\textcolor{blue}{38.24}$/$\textcolor{blue}{0.9614}$ & $38.13$/$0.9610$ & $\textcolor{red}{38.25}$/$\textcolor{red}{0.9614}$\\
			& $\times 3$ & -/- & $34.38$/$0.9262$ & -/- & $34.65$/$0.9282$ & $34.66$/$0.9280$ & $34.71$/$0.9296$ & $\textcolor{blue}{34.75}$/$\textcolor{blue}{0.9296}$ & $\textcolor{red}{34.83}$/$\textcolor{red}{0.9303}$\\
			& $\times 4$ & $32.02$/$0.8934$ & $32.07$/$0.8903$ & $32.47$/$0.8980$ & $32.46$/$0.8968$ & $32.50$/$0.8973$ & $32.47$/$0.8990$ & $\textcolor{blue}{32.57}$/$\textcolor{blue}{0.8991}$ & $\textcolor{red}{32.71}$/$\textcolor{red}{0.9008}$\\\hline\hline\vspace*{.03cm}
			\multirow{3}{*}{Set14} & $\times 2$ & -/- & $33.74$/$0.9170$ & $33.85$/$0.9190$ & $33.92$/$0.9195$ & $33.85$/$0.9198$ & $34.01$/$0.9212$ & $\textcolor{blue}{34.09}$/$\textcolor{blue}{0.9219}$ & $\textcolor{red}{34.15}$/$\textcolor{red}{0.9227}$\\
			& $\times 3$ & -/- & $30.34$/$0.8395$ & -/- & $30.52$/$0.8462$ & $30.44$/$0.8452$ & $30.57$/$\textcolor{blue}{0.8468}$ & $\textcolor{blue}{30.58}$/$0.8465$ & $\textcolor{red}{30.67}$/$\textcolor{red}{0.8481}$\\
			& $\times 4$ & $28.50$/$0.7782$ & $28.60$/$0.7751$ & $\textcolor{blue}{28.82}$/$0.7860$ & $28.80$/$\textcolor{blue}{0.7876}$ & $28.72$/$0.7857$ & $28.81$/$0.7871$ & $28.79$/$0.7867$ & $\textcolor{red}{28.93}$/$\textcolor{red}{0.7894}$\\\hline\hline\vspace*{.03cm}
			\multirow{3}{*}{B100} & $\times 2$ & -/- & $32.23$/$0.9013$ & $32.27$/$0.9000$ & $32.32$/$0.9013$ & $32.29$/$0.9007$ & $\textcolor{blue}{32.34}$/$\textcolor{blue}{0.9017}$ & $32.32$/$0.9014$ & $\textcolor{red}{32.38}$/$\textcolor{red}{0.9020}$\\
			& $\times 3$ & -/- & $29.08$/$0.8041$ & -/- & $29.25$/$0.8093$ & $29.25$/$0.8091$ & $\textcolor{blue}{29.26}$/$\textcolor{blue}{0.8093}$ & $\textcolor{blue}{29.26}$/$0.8091$ & $\textcolor{red}{29.32}$/$\textcolor{red}{0.8103}$\\
			& $\times 4$ & $27.53$/$0.7337$ & $27.52$/$0.7273$ & $27.72$/$0.7400$ & $27.71$/$\textcolor{blue}{0.7420}$ & $27.72$/$0.7418$ & $27.72$/$0.7419$ & $\textcolor{blue}{27.73}$/$0.7414$ & $\textcolor{red}{27.80}$/$\textcolor{red}{0.7430}$\\\hline\hline\vspace*{.03cm}
			\multirow{3}{*}{Urban100} & $\times 2$ & -/- & $32.22$/$0.9326$ & $32.55$/$0.9324$ & $\textcolor{blue}{32.93}$/$0.9351$ & $32.84$/$0.9347$ & $32.89$/$0.9353$ & $\textcolor{blue}{32.93}$/$\textcolor{blue}{0.9355}$ & $\textcolor{red}{33.14}$/$\textcolor{red}{0.9370}$\\
			& $\times 3$ & -/- & $28.08$/$0.8554$ & -/- & $28.80$/$0.8653$ & $28.79$/$0.8655$ & $28.80$/$0.8653$ & $\textcolor{blue}{28.87}$/$\textcolor{blue}{0.8667}$ & $\textcolor{red}{29.09}$/$\textcolor{red}{0.8697}$\\
			& $\times 4$ & $26.05$/$0.7819$ & $26.04$/$0.7896$ & $26.38$/$0.7946$ & $26.64$/$0.8033$ & $26.67$/$0.8041$ & $26.61$/$0.8028$ & $\textcolor{blue}{26.69}$/$\textcolor{blue}{0.8045}$ & $\textcolor{red}{26.90}$/$\textcolor{red}{0.8089}$\\\hline\hline\vspace*{.03cm}
			\multirow{3}{*}{Manga109} & $\times 2$ & -/- & $38.82$/$\textcolor{red}{0.9868}$ & $38.89$/$0.9775$ & $39.10$/$0.9773$ & $38.96$/$0.9769$ & $\textcolor{blue}{39.18}$/$0.9780$ & $39.07$/$0.9778$ & $\textcolor{red}{39.28}$/$\textcolor{blue}{0.9784}$\\
			& $\times 3$ & -/- & $33.44$/$0.9427$ & -/- & $34.17$/$0.9476$ & $34.17$/$0.9473$ & $34.13$/$\textcolor{blue}{0.9484}$ & $\textcolor{blue}{34.17}$/$0.9481$ & $\textcolor{red}{34.52}$/$\textcolor{red}{0.9498}$\\
			& $\times 4$ & -/- & $30.17$/$0.9034$ & $30.91$/$0.9137$ & $31.02$/$0.9148$ & $31.11$/$0.9148$ & $31.00$/$0.9151$ & $\textcolor{blue}{31.16}$/$\textcolor{blue}{0.9154}$ & $\textcolor{red}{31.52}$/$\textcolor{red}{0.9187}$
			\\\hline
		\end{tabular}
		\vspace*{-.1cm}
		\caption{\label{tab:psnr2}Average PSNR/SSIMs of \textbf{Post-upsampling} models for scale factor $\times2$, $\times3$ and $\times4$ on datasets Set5, Set14, BSD100, Urban100 and Manga109. \textcolor{red}{Red} color indicates the best performance and \textcolor{blue}{blue} color indicates the second best performance.}
		\vspace*{-.3cm}
	\end{table*}
}

\begin{figure*}
	\renewcommand{\thesubfigure}{} 
	\vspace*{-.2cm}
	\centering
	\includegraphics[width=\linewidth, keepaspectratio]{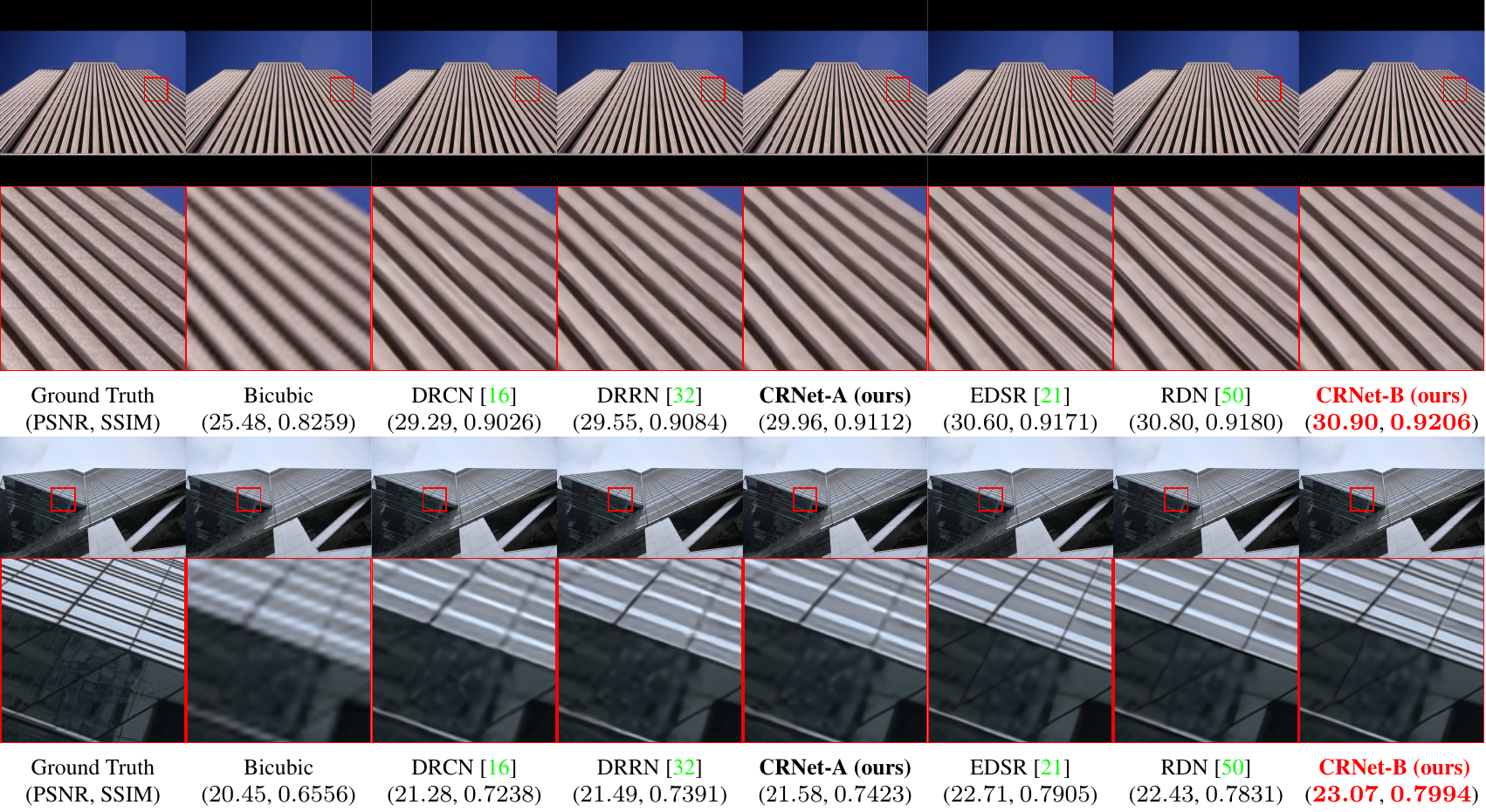}
	\caption{\label{fig:visual}SR results of ``img016'' and ``img059'' from \textbf{Urban100} with scale factor $\times 4$. \textcolor{red}{Red} indicates the best performance.}
	\vspace*{-.33cm}
	%
\end{figure*}

\subsection{Datasets and metrics}

\textbf{Training Set} By following \cite{VDSR2016, DRRN2017}, the training set of CRNet-A consists of \textit{291} images, where $91$ of these images are from Yang \etal~\cite{JianChaoYang2010} with the addition of $200$ images from Berkeley Segmentation Dataset \cite{BSD100:2001}. For CRNet-B, $800$ training images of \textit{DIV2K} \cite{NTIRE2017} are used for training.

\textbf{Testing Set} During testing, \textit{Set5} \cite{Set5:2012}, \textit{Set14} \cite{Set14:2010}, \textit{B100} \cite{BSD100:2001} and \textit{Urban100} \cite{Urban100:2015} are employed. As recent post-upsampling methods \cite{EDSR2017, MSRN2018, DDBPN2018, RDN2018} also evaluate their performance on \textit{Manga109} \cite{Manga109:2017}, so does CRNet-B.

\textbf{Metrics} Both PSNR and SSIM \cite{SSIM2004} on Y channel (\ie, luminance) of transformed YCbCr space are calculated for evaluation.

\subsection{Implementation details}
\label{subsec:details}

\textbf{CRNet-A} Data augmentation and
scale augmentation \cite{VDSR2016, DRCN2016, DRRN2017} are used for training a single model for all different scales ($\times 2$, $\times 3$ and $\times 4$). Every convolution layer in CRNet-A contains $128$ filters ($n_0 = 128$) of size $3\times 3$ while $\bm{W}_l$ and $\bm{S}$ have $256$ filters ($m_0 = 256$). The network is optimized using SGD. The learning rate is initially set to $0.1$ and then decreased by a factor of $10$ every $10$ epochs. L2 loss is used for CRNet-A, and we train a total of $35$ epochs.

\textbf{CRNet-B} Every weight layer in CRNet-B has $64$ filters ($n_0 = 64$) with the size of $3\times 3$ except $\bm{W}_l$ and $\bm{S}$ have $1,024$ filters ($m_0 = 1,024$). CRNet-B is updated using Adam \cite{Adam2014}. The initial learning rate is $10^{-4}$ and halved every $200$ epochs. We train CRNet-B for $800$ epochs. Unlike CRNet-A, CRNet-B is trained using L1 loss for better convergence speed.

\textbf{Recursion} We choose $K=25$ in both of our models. We implement our models using the PyTorch \cite{pytorch2017} framework with NVIDIA Titan Xp. It takes approximately $4.5$ days to train CRNet-A, and $15$ days to train CRNet-B.

\subsection{Comparison with CSC-SR}

We first compare our proposed models with the existing CSC based image SR method,  \ie, CSC-SR \cite{CSC-SR2015}. Since CSC-SR utilizes LR images as input image, it can be considered as a post-upsampling method, thus CRNet-B is used for comparison. Tab.~\ref{tab:compare:cscsr} presents that our CRNet-B clearly outperforms CSC-SR by a large margin.

\subsection{Comparison with State of the Arts}
\label{subsec:start-of-the-arts}

We now compare the proposed models with other state-of-the-arts in recent years. We compare CRNet-A with pre-upsampling models (\ie, SRCNN \cite{SRCNN2016}, RED30 \cite{REDNet2016}, VDSR \cite{VDSR2016}, DRCN \cite{DRCN2016}, DRRN \cite{DRRN2017}, MemNet \cite{MemNet2017}) while CRNet-B with post-upsampling architectures (\ie, SRDenseNet \cite{SRDenseNet2017}, MSRN \cite{MSRN2018}, D-DBPN \cite{DDBPN2018}, EDSR/MDSR \cite{EDSR2017}, RDN \cite{RDN2018}). Similar to \cite{EDSR2017,RDN2018}, self-ensemble strategy \cite{EDSR2017} is also adopted to further improve the performance of CRNet-B, and we denote the self-ensembled version as CRNet-B+.

Tab.~\ref{tab:psnr1} and Tab.~\ref{tab:psnr2} show the quantitative comparisons on the benchmark testing sets. Both of our models achieve superior performance against the state-of-the-arts, which indicates the effectiveness of our models. Qualitative results are provided in Fig.~\ref{fig:visual}. Our methods tend to produce shaper edges and more correct textures, while other images may be blurred or distorted. More visual comparisons are available in the supplementary material.

Fig.~\ref{fig:psnr:params} shows the performance versus the number of parameters, our CRNet-B and CRNet-B+ achieve better results with fewer parameters than EDSR \cite{EDSR2017} and RDN \cite{RDN2018}. It's worth noting that EDSR/MDSR and RDN are far deeper than CRNet-B (\eg, $169$ \vs $36$), but CRNet-B is quite wider ($\bm{W}_l$ and $\bm{S}$ have $1,024$ filters). As reported in \cite{EDSR2017}, when increasing the number of filters to a certain level, \eg, $256$, the training procedure of EDSR (for $\times 2$) without residual scaling \cite{Incepv4-2016, EDSR2017}  is numerically unstable, as shown in Fig.~\ref{sub:EDSR:loss}. However, CRNet-B is relieved from  the residual scaling trick. The training loss of CRNet-B is depicted in Fig.~\ref{sub:RLCSCB:loss}, it converges fast at the begining, then keeps decreasing and finally fluctuates at a certain range.


\subsection{Parameter Study}

The key parameters in both of our models are the number of filters ($n_0, m_0$) and recursions $K$. 

\begin{figure}
	\centering
	\renewcommand{\thesubfigure}{(a)} 
	\vspace*{-.2cm}
	\subfloat[Pre-upsampling models]{\label{sub:bicubic:params}\includegraphics[%
		width=0.466\linewidth, keepaspectratio]{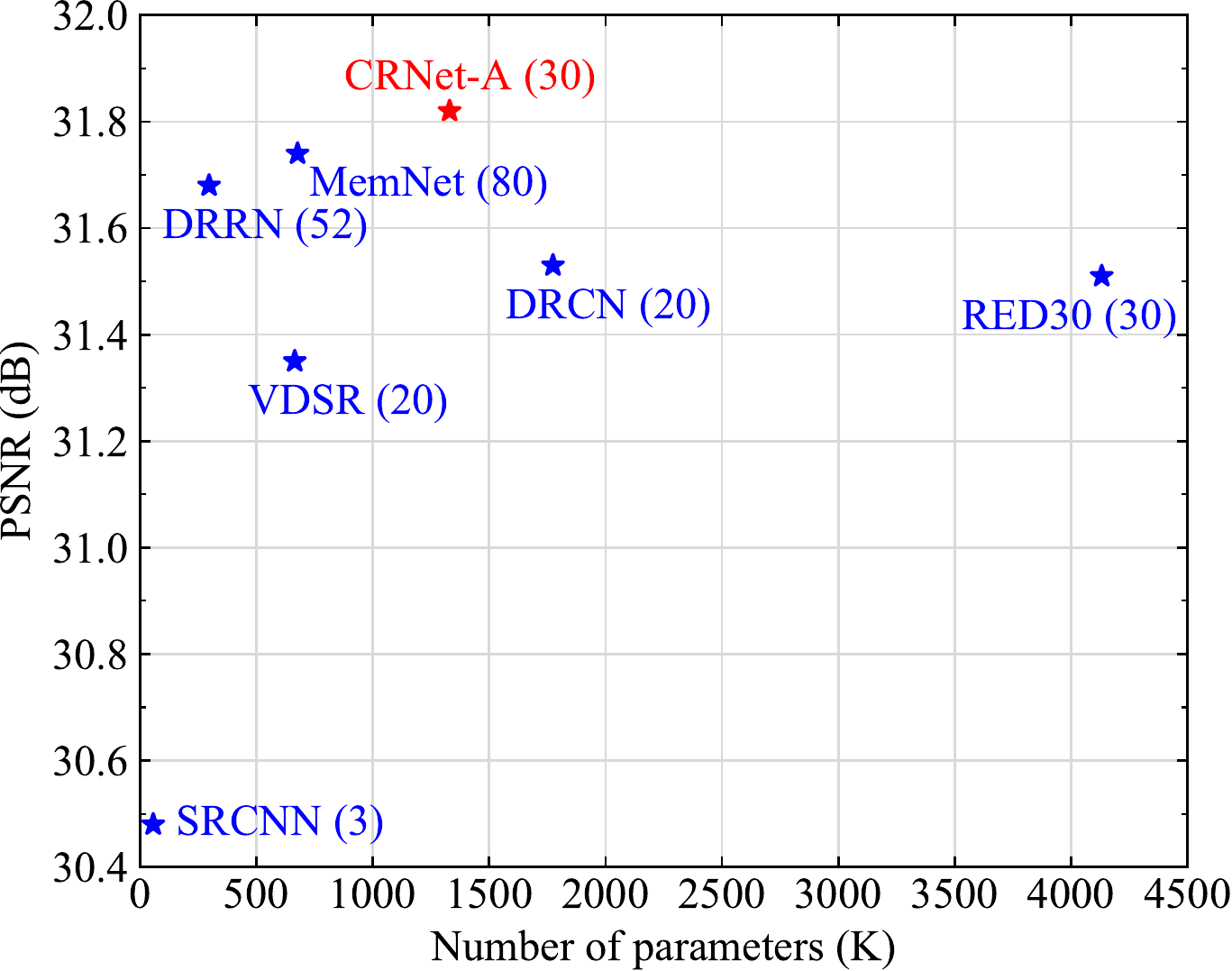}}\hspace*{.03\linewidth}
	\renewcommand{\thesubfigure}{(b)}
	\subfloat[Post-upsampling models]{\label{sub:nobicubic:params}\includegraphics[%
		width=0.46\linewidth, keepaspectratio]{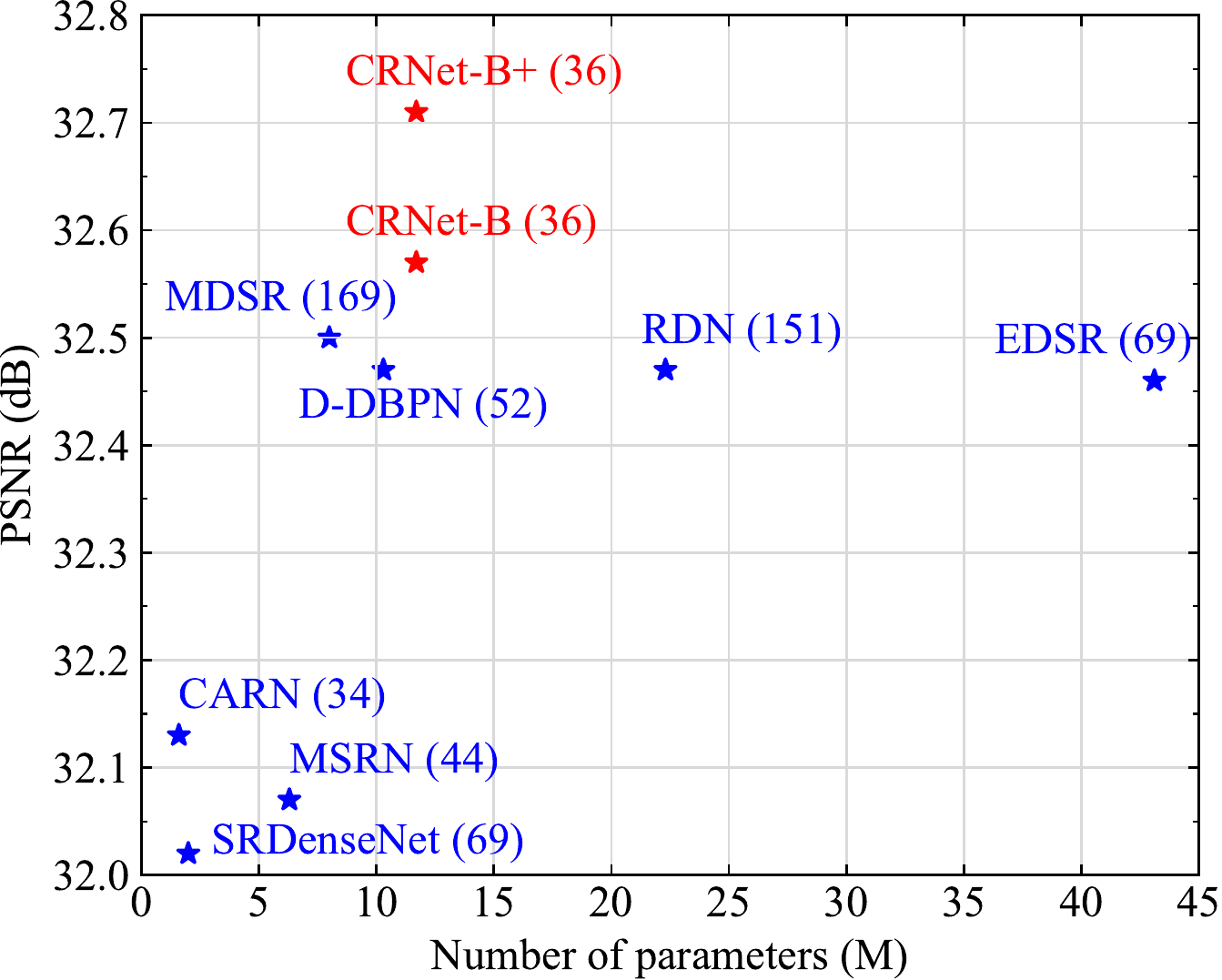}}
	\vspace*{-.2cm}
	\caption{\label{fig:psnr:params} PSNR of recent state-of-the-arts versus the number of parameters for scale factor $\times 4$ on Set5. The number of layers are marked in the parentheses.}
	
	\centering
	\renewcommand{\thesubfigure}{(a)} 
	\vspace*{-.2cm}
	\subfloat[EDSR for $\times 2$]{\label{sub:EDSR:loss}\includegraphics[%
		width=0.461\linewidth, keepaspectratio]{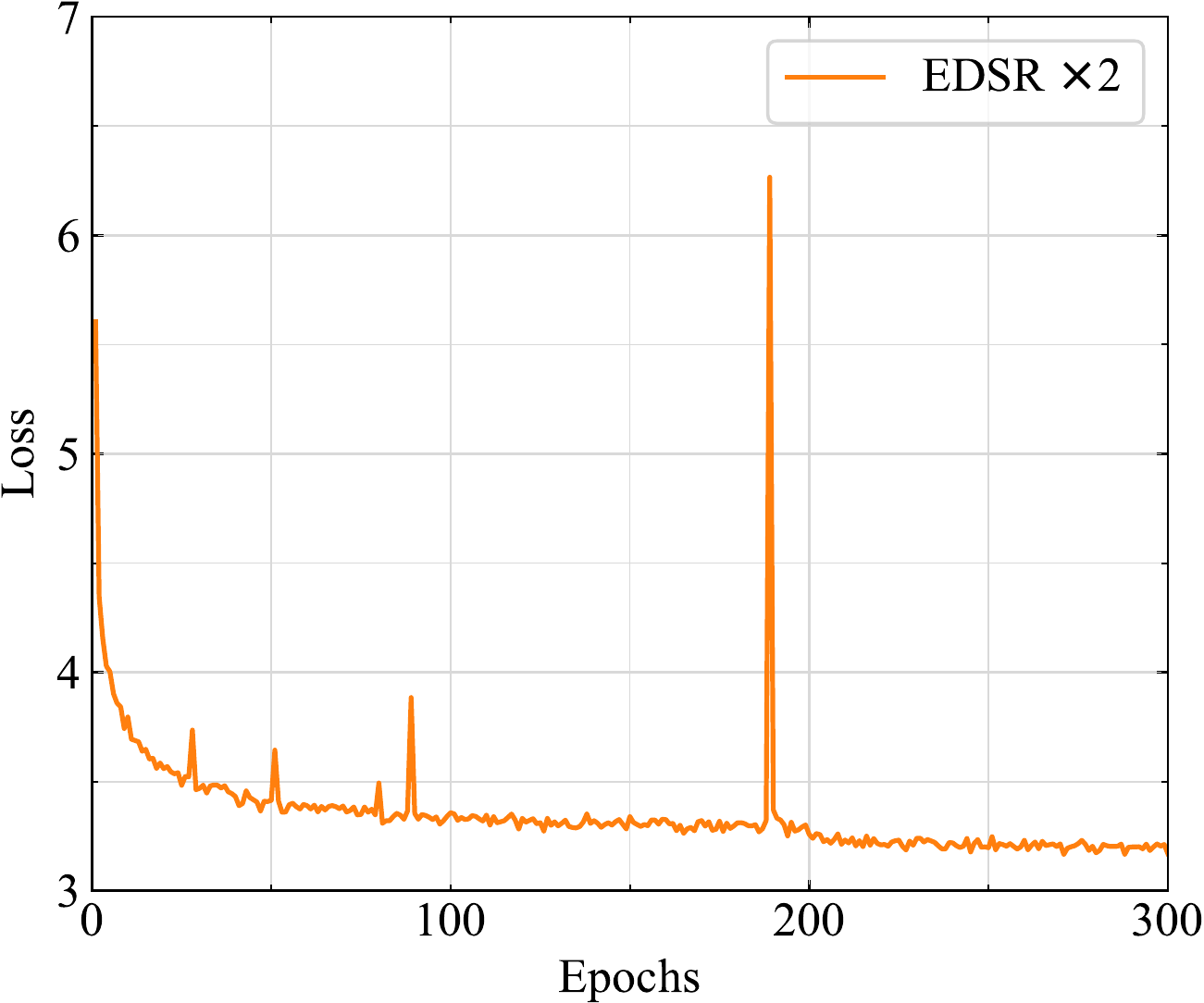}}\hspace*{.03\linewidth}
	\renewcommand{\thesubfigure}{(b)}
	\subfloat[CRNet-B for all scales]{\label{sub:RLCSCB:loss}\includegraphics[%
		width=0.47\linewidth, keepaspectratio]{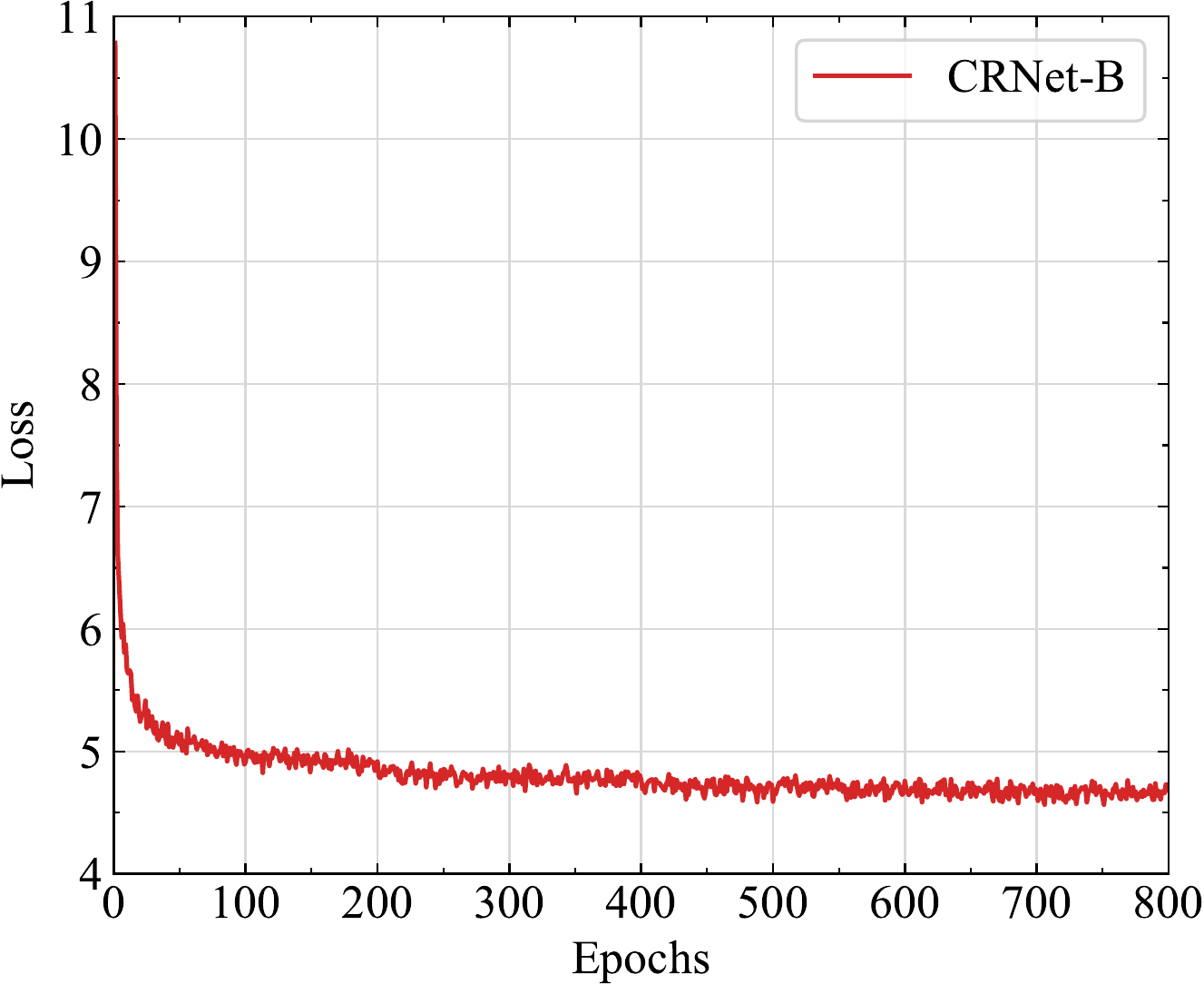}}
	\vspace*{-.2cm}
	\caption{\label{fig:RLCSCB:lines} Training loss of  EDSR ($\times 2$) without residual scaling and CRNet-B (for all scales).}
	%
	%
	\centering
	\renewcommand{\thesubfigure}{(a)} 
	\vspace*{-.2cm}
	\subfloat[CRNet-A]{\label{sub:RLCSCA:filters}\includegraphics[%
		width=0.46\linewidth, keepaspectratio]{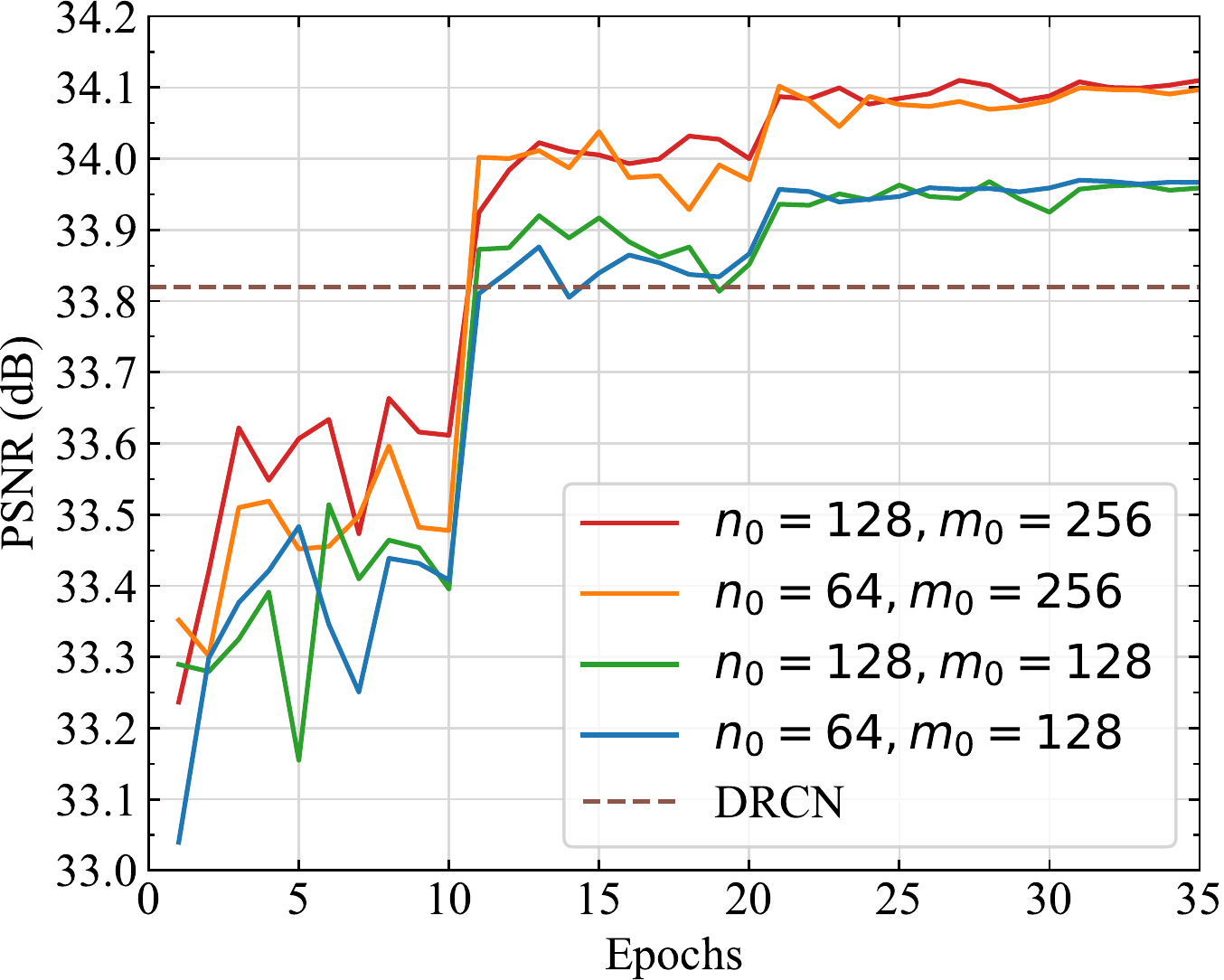}}\hspace*{.03\linewidth}
	\renewcommand{\thesubfigure}{(b)}
	\subfloat[CRNet-B]{\label{sub:RLCSCB:params}\includegraphics[%
		width=0.471\linewidth, keepaspectratio]{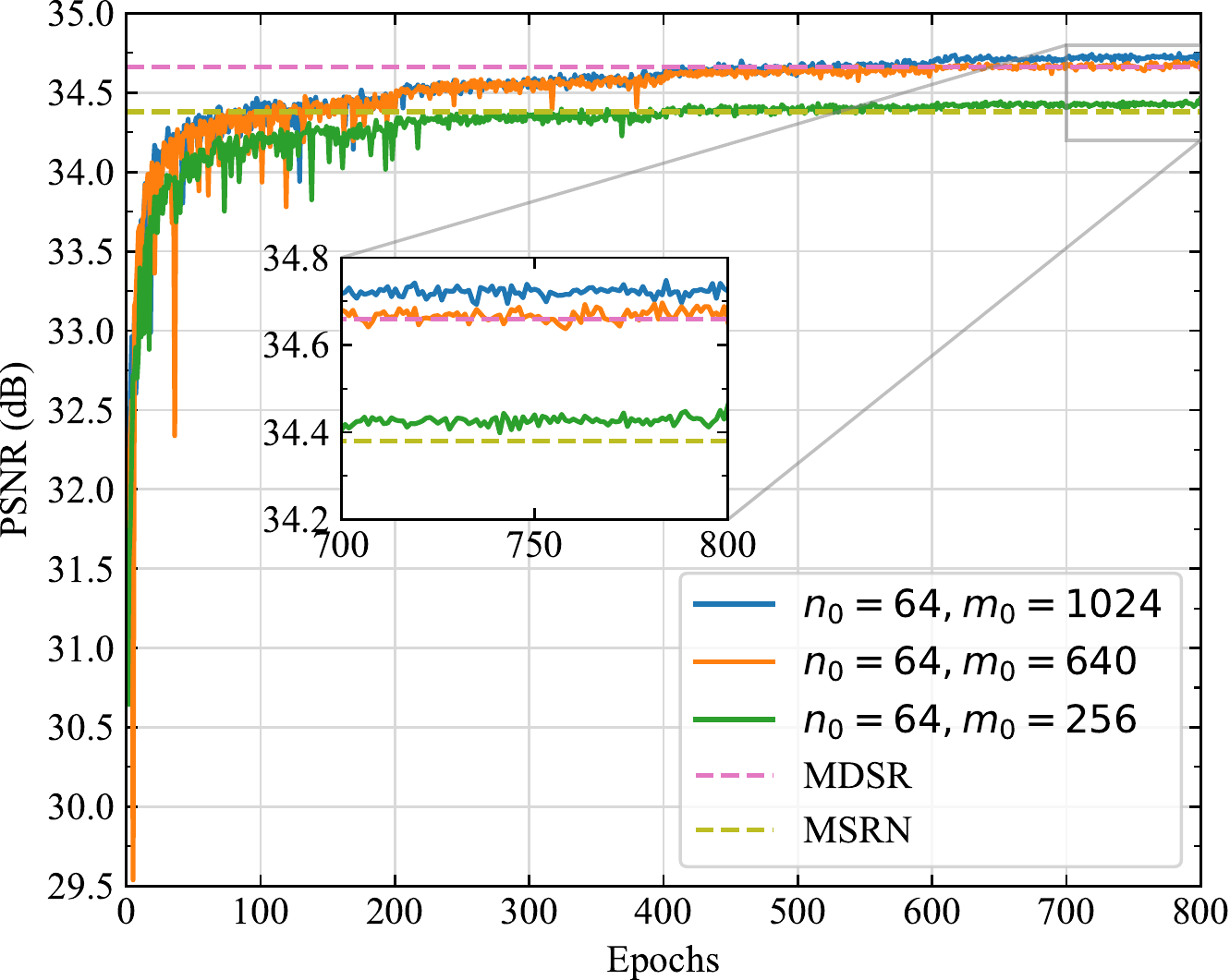}}
	\vspace*{-.2cm}
	\caption{\label{fig:num:filters} PSNR of proposed models versus different number of filters on Set5 with scale factor $\times 3$.}
	
	\centering
	\renewcommand{\thesubfigure}{(a)} 
	\vspace*{-.2cm}
	\subfloat[CRNet-A]{\label{sub:RLCSCA:K}\includegraphics[%
		width=0.46\linewidth, keepaspectratio]{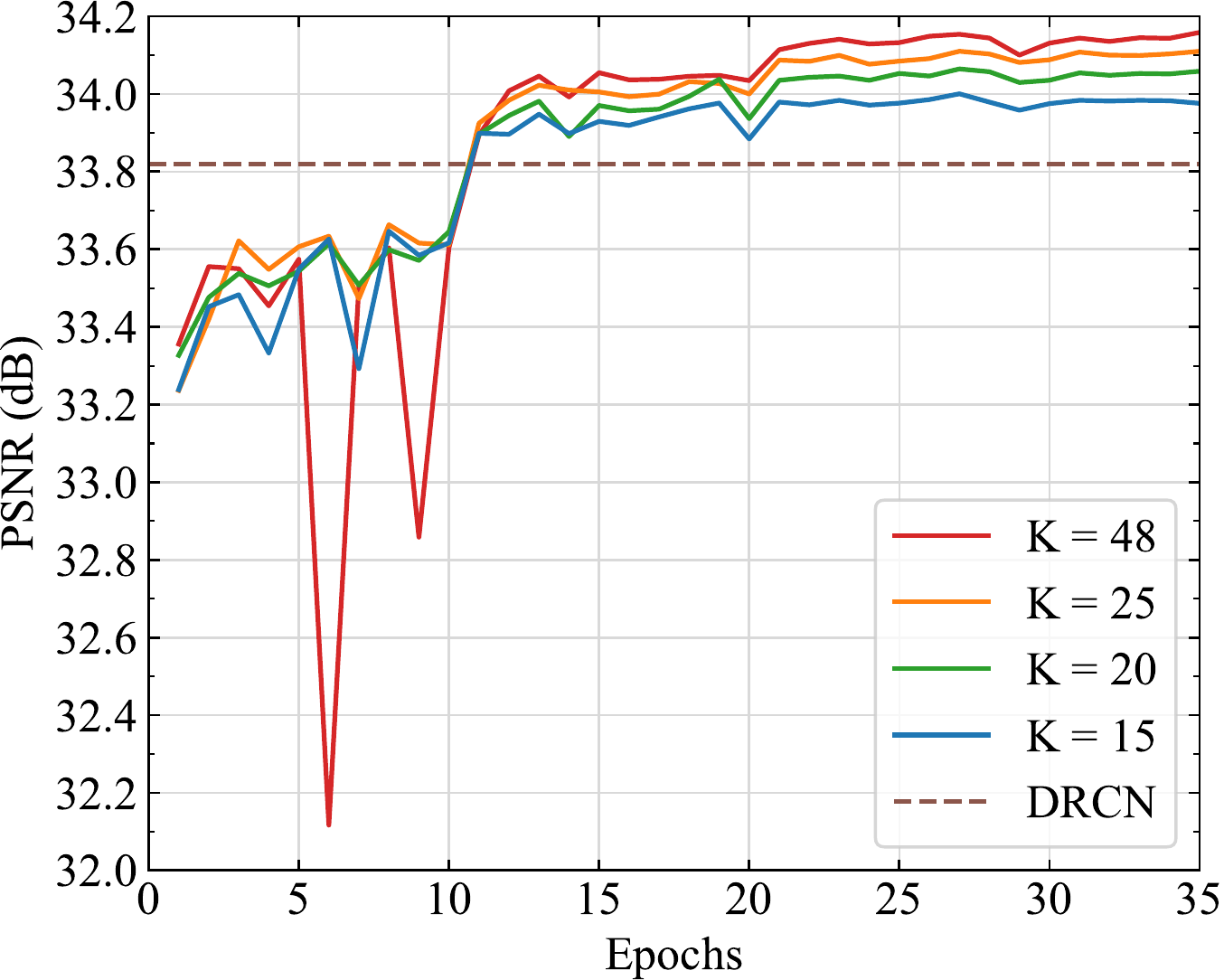}}\hspace*{.03\linewidth}
	\renewcommand{\thesubfigure}{(b)}
	\subfloat[CRNet-B]{\label{sub:RLCSCB:K}\includegraphics[%
		width=0.46\linewidth, keepaspectratio]{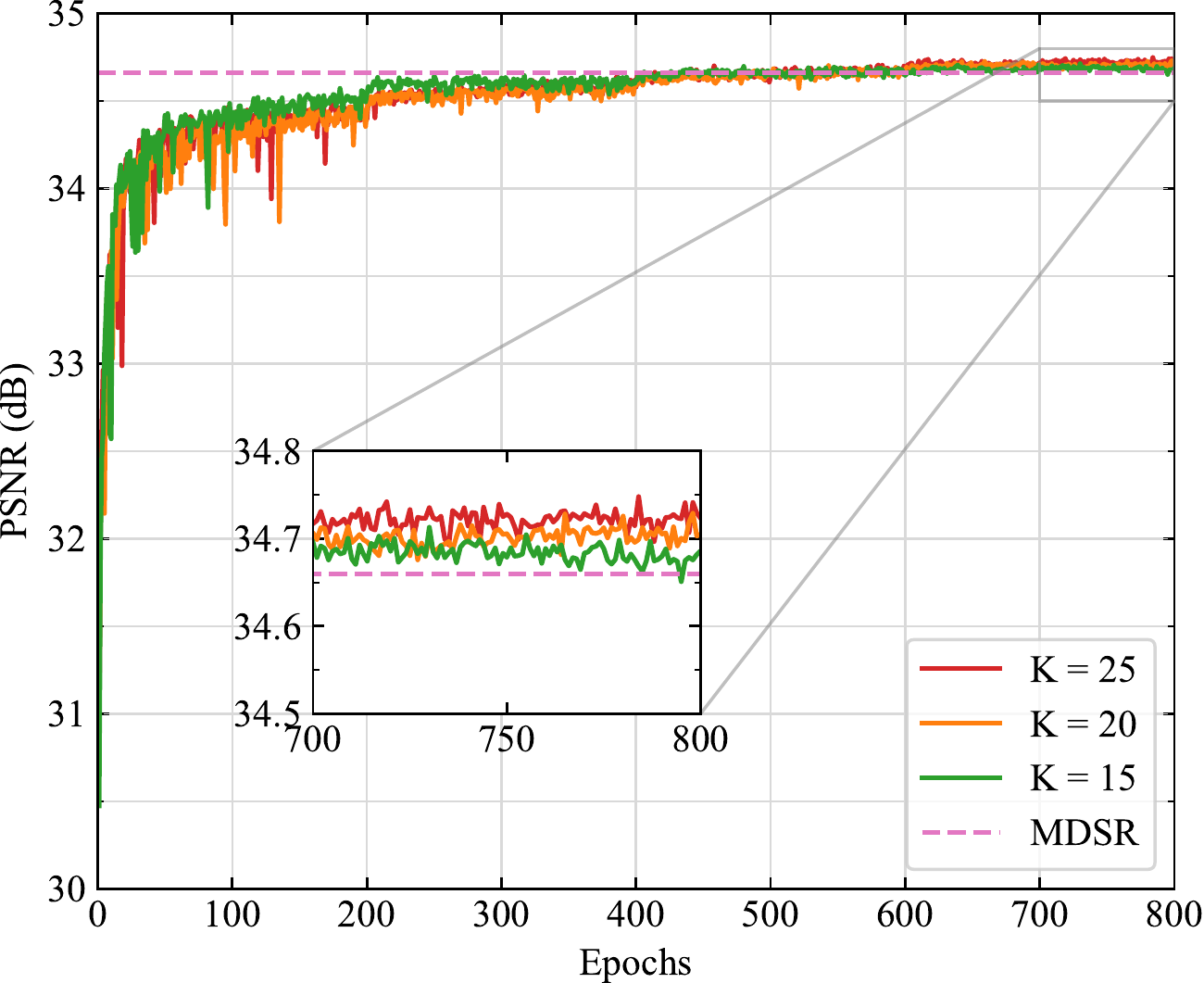}}
	\vspace*{-.2cm}
	\caption{\label{fig:num:recursions} PSNR of proposed models versus different number of recursions on Set5 with scale factor $\times 3$.}
	\vspace*{-.5cm}
\end{figure}

\textbf{Number of Filters} We set $n_0 = 128, m_0 = 256, K=25$ for CRNet-A as stated in Section~\ref{subsec:details}. In Fig.~\ref{sub:RLCSCA:filters}, CRNet-A with different number of filters are tested (DRCN \cite{DRCN2016} is used for reference). We find that even $n_0$ is decreased from $128$ to $64$, the performance is not affected greatly. On the other hand, if we decrease $m_0$ from $256$ to $128$, the performance would suffer an obvious drop, but still better than DRCN \cite{DRCN2016}. Based on these observations, we set the parameters of CRNet-B by making $m_0$ larger and $n_0$ smaller for the trade off between model size and performance. Specifically, we use $n_0 = 64, m_0 = 1024, K=25$ for CRNet-B. As shown in Fig.~\ref{sub:RLCSCB:params}, the performance of CRNet-B can be significantly boosted with larger $m_0$ (MDSR \cite{EDSR2017} and MSRN \cite{MSRN2018} are used for reference). Even with small $m_0$, \ie, $256$, CRNet-B still outperforms MSRN \cite{MSRN2018} with fewer parameters ($2.0$M \vs $6.1$M).

\textbf{Number of Recursions} We also have trained and tested CRNet-A with $15$, $20$, $25$, $48$ recursions, so the depth of the these models are $20$, $25$, $30$, $53$ respectively. The results are presented in Fig.~\ref{sub:RLCSCA:K}.  It's clear that CRNet-A with $20$ layers still outperforms DRCN with the same depth and increasing $K$ can promote the final performance. The results of using different recursions in CRNet-B are shown in Fig.~\ref{sub:RLCSCB:K}, which demonstrate that more recursions facilitate the performance improved.

\section{Discussions}
\begin{figure}
	\centering
	\renewcommand{\thesubfigure}{(a)}
	\subfloat[DRRN]{\includegraphics[width=.24\linewidth, keepaspectratio]{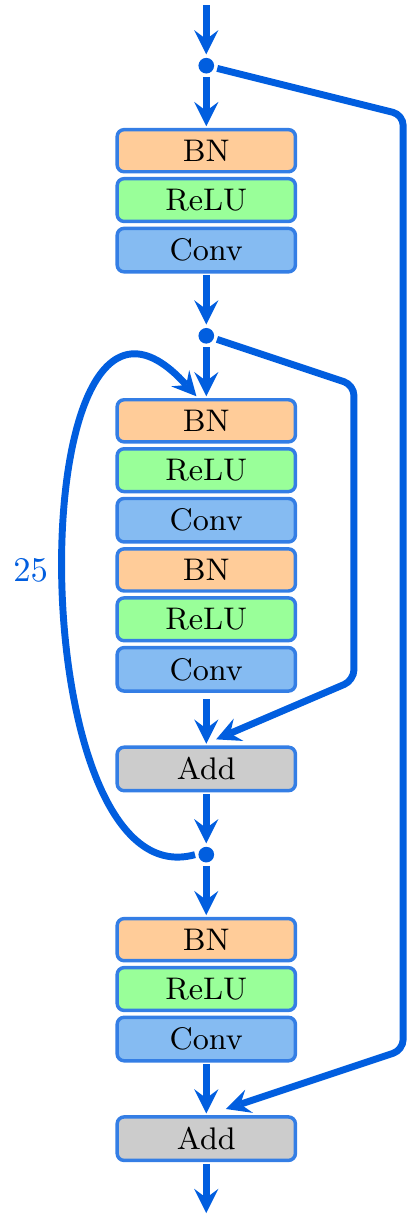}}
	\renewcommand{\thesubfigure}{(b)}
	\subfloat[SCN]{\includegraphics[width=.25\linewidth, keepaspectratio]{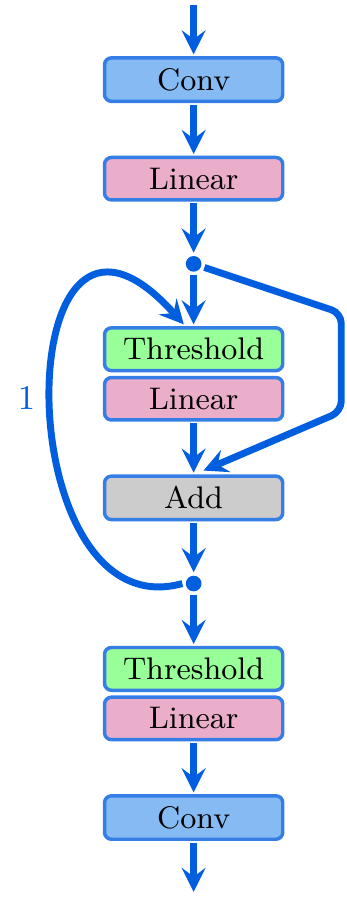}}
	\renewcommand{\thesubfigure}{(c)}
	\subfloat[DRCN]{\label{subfig:drcn}\includegraphics[width=.233\linewidth, keepaspectratio]{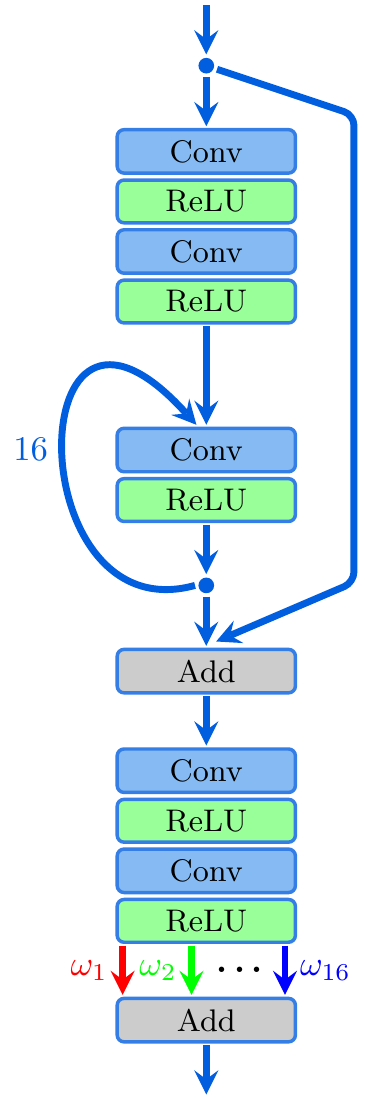}}
	\renewcommand{\thesubfigure}{(d)}
	\subfloat[CRNet-A]{\includegraphics[width=.23\linewidth, keepaspectratio]{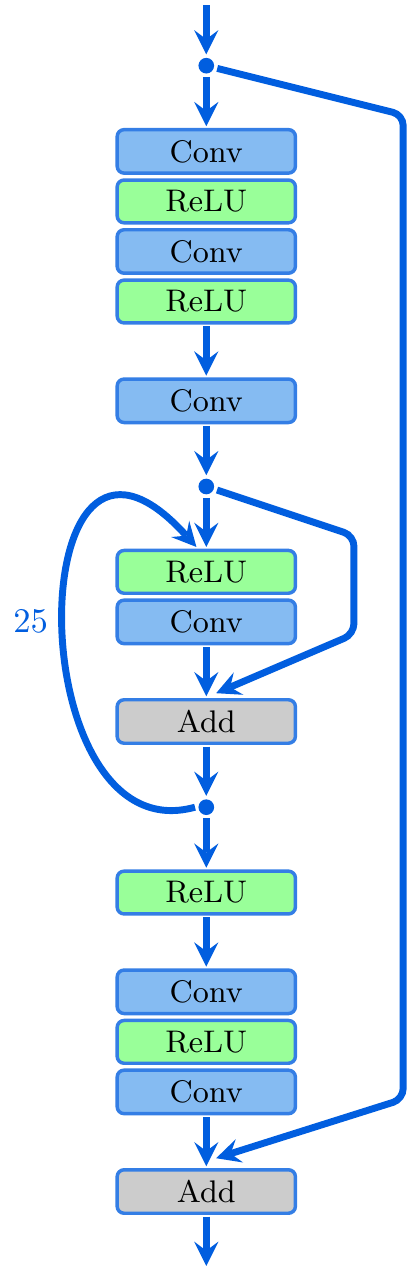}}
	\caption{\label{fig:comparisons} Simplified network structures of (a) DRRN \cite{DRRN2017}, (b) SCN \cite{SCN2015}, (c) DRCN \cite{DRCN2016}, (d) our model CRNet-A.}
	\vspace*{-.3cm}
\end{figure}


We discuss the differences between our proposed models and several recent CNN models for SR \textit{with recursive learning strategy}, \ie, DRRN \cite{DRRN2017}, SCN \cite{SCN2015} and DRCN \cite{DRCN2016}. Due to the fact that CRNet-B is an extension of CRNet-A, \ie, the main part of CRNet-B has the same structure as CRNet-A, so we use CRNet-A here for comparison. The simplified structures of these models are shown in Fig.~\ref{fig:comparisons}, where the digits on the left of the recursion line represent the number of recursions.

\noindent\textbf{Difference to DRRN}. The main part of DRRN \cite{DRRN2017} is the recursive block structure, where several residual units with BN layers are stacked. On the other hand, guided by \eqref{equ:cista}, CRNet-A contains no BN layers. Coinciding with EDSR/MDSR \cite{EDSR2017}, by normalizing features, BN layers get rid of range flexibility from networks. Furthermore, BN consumes much amount of GPU memory and increases computational complexity. Experimental results on benchmark datasets under common-used assessments demonstrate the superiority of CRNet-A.

\noindent\textbf{Difference to SCN}. There are two main differences between CRNet-A and SCN \cite{SCN2015}: CISTA block and residual learning. Specifically, CRNet-A takes consistency constraint into consideration with the help of CISTA block, while SCN uses linear layers and ignores the information from the consistency prior. On the other hand, CRNet-A adopts residual learning, which is a powerful tool for training deeper networks. CRNet-A ($30$ layers) is much deeper than SCN ($5$ layers). As indicated in \cite{VDSR2016}, a deeper network has larger receptive fileds, so more contextual information in an image would be utilized to infer high-frequency details. In Fig.~\ref{sub:RLCSCA:K}, we show that more recursions, \eg,  $48$, can be used to achieve better performance.

\noindent\textbf{Difference to DRCN}. CRNet-A differs with DRCN \cite{DRCN2016} in two aspects: recursive block and training techniques. In the recursive block, both local residual learning \cite{DRRN2017} and  pre-activation \cite{he2016identity, DRRN2017} are utilized in CRNet-A, which are demonstrated to be effective in \cite{DRRN2017}. 
As for training techniques, DRCN is not easy to train, so recursive-supervision is introduced to facilitate the network to converge. Moreover, an ensemble strategy (in Fig.~\ref{subfig:drcn}, the final output is the weighted average of all intermediate predictions) is used to further improve the performance. CRNet-A is relieved from these techniques and can be easily trained with more recursions.

\section{Conclusions}

In this work, we propose two effective CSC based image SR models, \ie, CRNet-A and CRNet-B, for pre-/post-upsampling SR, respectively. By combining the merits of CSC and CNN, we achieve superior performance against recent state-of-the-arts. Furthermore, our framework and CISTA block are expected to be applicable in various CSC based tasks,  though in this paper we focus on CSC based image SR.  


\newpage

{\small
\bibliographystyle{ieee}
\bibliography{reference}
}

\end{document}